\documentclass[sigconf, 9pt]{acmart}

\usepackage{enumitem}
\usepackage{graphicx}
\usepackage{caption}
\usepackage{subcaption}
\usepackage{comment}
\usepackage{pifont}
\usepackage{algorithm}
\usepackage{algpseudocode}
\usepackage{amsmath}
\usepackage{xspace}
\usepackage{makecell}

\graphicspath{ {./figures/} }

\usepackage[normalem]{ulem}
\AtBeginDocument{%
  }

\definecolor{iris}{rgb}{0.35, 0.31, 0.81}

\newcommand{\camready}[1]{\textcolor{black}{#1}}

\newcommand{\takeaways}{\noindent\textit{\textbf{Takeaways: }}}

\newcommand{\sys}{\emph{Aceso}\xspace}

\newcommand{\cmark}{\ding{51}}%
\newcommand{\xmark}{\ding{55}}%

\newenvironment{tightitemize}%
	 {\begin{list}{$\bullet$}{%
 		\setlength{\leftmargin}{10pt}
        \setlength{\itemsep}{0pt}%
        \setlength{\parsep}{0pt}%
        \setlength{\topsep}{0pt}%
        \setlength{\parskip}{0pt}%
        }%
 }%
{\end{list}}

\newcounter{tecounter}
\newenvironment{tightenumerate}%
 {\begin{list}{\arabic{tecounter}.}{%
 		\usecounter{tecounter}
 		\setlength{\leftmargin}{10pt}
        \setlength{\itemsep}{0pt}%
        \setlength{\parsep}{0pt}%
        \setlength{\topsep}{0pt}%
        \setlength{\parskip}{0pt}%
        }%
 }%
{\end{list}}%

\copyrightyear{2026}
\acmYear{2026}
\setcopyright{cc}
\setcctype{by}
\acmConference[Middleware '26]{27th International Middleware Conference}{December 14--18, 2026}{Tarragona, Spain}
\acmBooktitle{27th International Middleware Conference (Middleware '26), December 14--18, 2026, Tarragona, Spain}
\acmDOI{10.1145/3801927.3858642}
\acmISBN{979-8-4007-2621-7/2026/12}

\begin{document}

\title{Aceso: Latency-Aware Geo-Distribution of Microservices for Cost-Effective Carbon Reduction}

\author{Georgia Christofidi}
\orcid{0009-0007-8422-1167}
\affiliation{%
  \institution{IMDEA Software Institute}
  \institution{Universidad Politécnica de Madrid}
  \city{Madrid}
  \country{Spain}}
\email{georgia.christofidi@imdea.org}
  
\author{Francisco Álvarez-Terribas}
\orcid{0000-0001-8462-0175}
\affiliation{%
  \institution{Telefonica Research}
  \institution{Universidad Pompeu Fabra}
  \city{Barcelona}
  \country{Spain}}
\email{francisco.alvarezterribas@telefonica.com}
  
\author{Ioannis Roumpos}
\orcid{0009-0006-4606-7448}
\affiliation{%
  \institution{IMDEA Software Institute}
  \institution{Universidad Politécnica de Madrid}
  \city{Madrid}
  \country{Spain}}
\email{giannis.roumpos@imdea.org}
  
\author{Nicolas Kourtellis}
\orcid{0000-0002-5674-1698}
\affiliation{%
  \institution{Keysight Technologies}
  \city{Barcelona}
  \country{Spain}}
\email{nkourtellis@gmail.com}
  
\author{Jesus Omaña Iglesias}
\orcid{0009-0002-5473-2170}
\affiliation{%
  \institution{Telefonica Research}
  \city{Barcelona}
  \country{Spain}}
\email{jesusalberto.omana@telefonica.com}

\author{Thaleia Dimitra Doudali}
\orcid{0000-0002-3197-839X}
\affiliation{%
  \institution{IMDEA Software Institute}
  \city{Madrid}
  \country{Spain}}
\email{thaleia.doudali@imdea.org}

\renewcommand{\shortauthors}{Georgia Christofidi et al.}

\begin{abstract}
Microservices are a dominant architecture in cloud computing, offering scalability and modularity, but also posing complex deployment challenges. As data centers contribute significantly to global carbon emissions, carbon-aware scheduling is a promising mitigation strategy. However, most existing solutions target batch, 
HPC, or serverless workloads and assume
global-scale infrastructure. Such an assumption does not hold for many  
regional small to medium-sized enterprises (SMEs) with microservice applications, 
due to geographic, cost, and data sovereignty constraints. 
In this paper, we present \textbf{\sys}, an Adaptive Carbon- and Efficiency-aware placement for microservices that considers carbon, cost, and latency constraints. \sys dynamically places microservices across geographically constrained regions using a scalable optimization 
with insight-based search space pruning techniques. Evaluation on a real-world deployment shows that \sys adapts to real-time changes in workload and carbon intensity and reduces carbon emissions by \camready{up to 37.4\%} and operational cost \camready{by up to 3.36\%, on average across days,} compared to a static deployment within a single country, while maintaining the application's latency targets. 
In this way, \sys enables carbon- and cost-aware microservice deployment for latency-sensitive applications in regionally limited infrastructures. 
\end{abstract}

\ccsdesc[500]{Computer systems organization~Cloud computing}
\ccsdesc[500]{Social and professional topics~Sustainability}

\keywords{Microservices, Cloud Computing, Geo-Distributed Systems, Geographical Workload Shifting, Carbon-Aware Computing, Sustainable Computing, Latency-Aware Optimization, Cost Optimization}

\maketitle

\section{Introduction}
\label{sec:intro}

The rapid growth of generative AI is driving unprecedented scale in cloud infrastructures~\cite{ai-agents, sustainable-ai-meta}, with microservices forming the backbone of large-scale AI systems~\cite{ml-scale-ms, ml-for-ms, ms-for-ml, ms-for-large-ai, ms-ai-guide, ms-modular-ai, ms-arch-ai}. 
While this combination accelerates innovation, 
it also significantly increases energy consumption and carbon emissions~\cite{on-the-limitations, Thangam2024datacenters, sustainable-ai-meta}. 
Over the past years, data centers worldwide contributed significantly to carbon emissions, and their impact is expected to increase further~\cite{power-modeling-2022, socc-serverless-carbon, Toonder2024}. This trend underscores the urgent need to optimize microservice deployment not only for performance and cost but also to minimize the carbon footprint, ensuring a more sustainable digital infrastructure~\cite{accountable-carbon-serverless,energy-and-ai-report-2025}.

Carbon-aware computing aims to reduce the environmental impact of cloud workloads by aligning their execution with periods or locations of lower carbon intensity. This can be achieved through {\it temporal shifting}, by delaying execution to times when clean energy is more available~\cite{ecolife, green-for-less-green, lets-wait-awhile, carbon-scaler, Spatiotemporal-online, on-the-limitations, carbonexplorer, bostandoost2024data, pause-resume}, or {\it spatial shifting}, by moving workloads to regions with greener electricity~\cite{caribou, carbonedge, Spatiotemporal-online, on-the-limitations}. Many existing carbon-aware solutions, as summarized in Table~\ref{tab:related_works}, assume access to global-scale cloud infrastructure~\cite{green-for-less-green, carbon-scaler, carbonexplorer, on-the-limitations, ecovisor}, where workloads can be flexibly shifted across regions or time zones. Notably, Meta's carbon-intelligent scheduling~\cite{carbonexplorer} targets {\it delay-tolerant batch workloads} (e.g., AI training, offline data processing), exploiting their flexibility to shift jobs across time and to distant low-carbon regions. Serverless approaches~\cite{caribou, ecolife} similarly exploit {\it stateless, relocatable} functions (e.g., AWS Lambda) to enable worldwide geo-distribution with minimal operational effort.

However, latency-critical microservice applications, such as ride-hailing services (e.g., Uber), remain underexplored due to their latency sensitivity, tight inter-service communication, and dynamic behavior. Their performance often depends on low-latency interactions between components~\cite{nautilus, ursa, seer}, 
and their modular and evolving nature requires placement strategies, both fine-grained and responsive to workload changes~\cite{deeprest, deepscaling}. While systems like Oakestra~\cite{oakestra}, Astraea~\cite{astraea}, Nautilus~\cite{nautilus}, and Imbres~\cite{imbres} offer advanced orchestration for microservices, they focus on performance and resource efficiency and not on their carbon impact. Applying carbon-aware strategies in microservice applications is particularly challenging. 
Temporal shifting 
will delay user-facing requests and degrade user satisfaction, so it is largely unavailable. Spatial shifting 
can reduce carbon, but it must be applied carefully: placing latency-critical services in distant regions can 
degrade 
latency.

Spatial shifting becomes even more challenging in {\it geographically constrained settings}, such as deployments by small and medium-sized enterprises (SMEs), which operate under limited budgets, constrained infrastructures, and primarily serve local or regional users. For example, local ride-hailing applications (e.g., Cabify~\cite{cabify} in Spain, itTaxi~\cite{ittaxi} in Italy, FreeNow~\cite{freenow} in Europe) handle highly localized traffic, so moving services to distant greener regions directly increases user-facing latency. Data sovereignty and privacy rules~\cite{data-survey-2024-peer-review, data-sovereignty-2024, gdpr} further limit migration, often requiring data to stay within specific jurisdictions (e.g., the European Union). SMEs often operate under strict cost budgets, which can constrain the extent to which such carbon reduction strategies can be realistically pursued; yet existing carbon-aware solutions rarely prioritize cost (Table~\ref{tab:related_works}). As a result, the assumptions of prior work regarding abundant geographic flexibility~\cite{carbonexplorer, Spatiotemporal-online}, negligible cost constraints~\cite{carbonedge, Lechowicz2025, hadidi2023keids}, and limited consideration of latency-sensitive applications~\cite{green-for-less-green,Lechowicz2025, carbon-scaler, pause-resume}, reduce their direct applicability.

At the same time, organizations of all sizes face mounting pressure to reduce their carbon footprint. Regulatory frameworks such as the European Green Deal~\cite{eugreendeal2019} and the European Climate Law~\cite{europeanclimatelaw2021}, along with sustainability reporting standards, are driving the adoption of greener practices. Financial instruments, including carbon taxes~\cite{carbon_tax_jrc2024}, emission trading schemes~\cite{eu_ets_2003}, and green procurement incentives~\cite{green_procurement_ec2025}, are turning carbon efficiency into a business imperative aligned with global sustainability goals, such as the UN Sustainable Development Goals (SDGs)~\cite{un_sdg_2015}. 
For SMEs with limited infrastructure flexibility, this is both a challenge 
and an opportunity to adopt carbon-aware, cost-efficient, regionally deployable solutions that support global climate goals. Therefore, this paper addresses the following question:

\vspace{0.05in}
\noindent\fbox{\parbox{0.95\linewidth}{
\emph{How can SMEs deploy microservice applications in a carbon-efficient and cost-effective way within geographically constrained infrastructures while maintaining performance?}
}}

To answer this question, we present \textbf{\sys}\footnote{In ancient Greek mythology, Aceso (pronounced Akesó) was the Greek goddess of well-being. Unlike her sister Panacea (Cure-All), she represented the process of curing rather than the cure itself.}: an Adaptive Carbon- and Efficiency-aware microservice placement system that jointly optimizes carbon emissions and monetary cost, while respecting latency constraints. \sys employs a scalable genetic algorithm that rapidly identifies near-optimal microservice placements across available regions that have heterogeneous carbon intensity.
Given the potentially exponential size of the search space, \sys leverages a key \textbf{insight} to effectively reduce it: 
it prioritizes the offloading of microservices that \emph{activate early} in the request execution path. 
Motivational analysis from a real-world deployment across diverse microservice applications (Social Network, Media Streaming, and Hotel Reservation) shows that placing early-activated services in regions with lower carbon intensity can yield substantial carbon reductions, while also improving cost-efficiency when such regions have lower operational costs. 
Overall, \sys achieves up to 37.4\% carbon emissions reduction and up to \camready{3.36\%} operational cost savings, on average across days, compared to a static, single-region deployment, while maintaining the application's latency targets.
Overall, \sys enables both carbon- and cost-aware microservice placement for latency-sensitive applications in regionally constrained infrastructures.
The specific {\bf paper contributions} are:
\begin{tightitemize}
    \item A characterization of carbon-reduction opportunities for microservices under geographic constraints, revealing that selectively offloading early-activated services to greener regions can lower emissions while maintaining performance (Section~\ref{sec:motivation}).
    \item The design of \sys, including a dynamic optimization strategy that adapts to workload patterns and carbon intensity using a scalable genetic algorithm (Section~\ref{sec:system}).
    \item A comprehensive evaluation of \sys, demonstrating significant reductions in carbon footprint and cost while maintaining application performance in a scalable and stable manner (Section~\ref{sec:evaluation}). 
    \item \sys is {\bf open sourced}\footnote{\url{https://github.com/muse-research-lab/aceso}} alongside a simulator for large-scale geo-distributed microservices, facilitating community adoption and reproducibility. 
\end{tightitemize}

\section{Background}
\label{sec:back}

\begin{figure}[t]
    \centering

    \begin{subfigure}{0.49\columnwidth}
        \centering
        \includegraphics[width=\columnwidth]{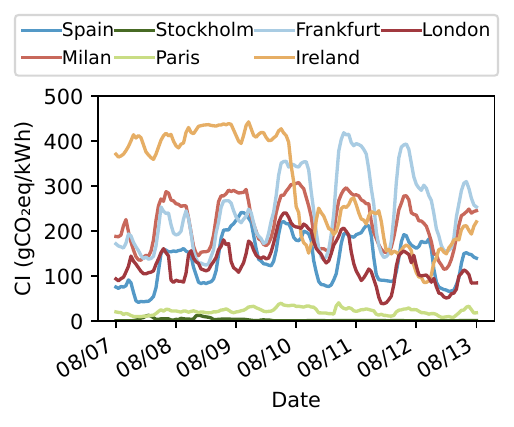}
        \vspace{-20pt}
        \caption{Europe}
        \label{fig:EU_CI_week}
    \end{subfigure}
    \hfill
    \begin{subfigure}{0.49\columnwidth}
        \centering
        \includegraphics[width=\columnwidth]{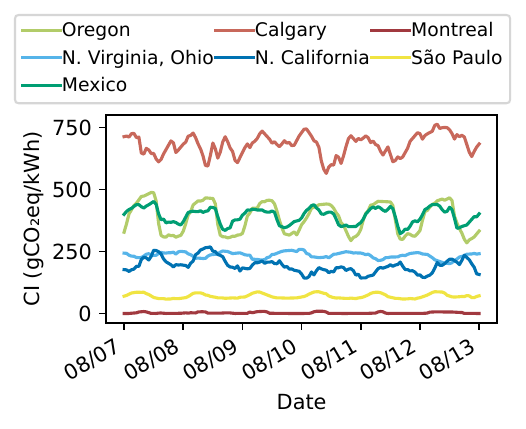}
        \vspace{-20pt}
        \caption{South \& North America}
        \label{fig:AM_CI_week}
    \end{subfigure}

    \begin{subfigure}{0.49\columnwidth}
        \centering
        \includegraphics[width=\columnwidth]{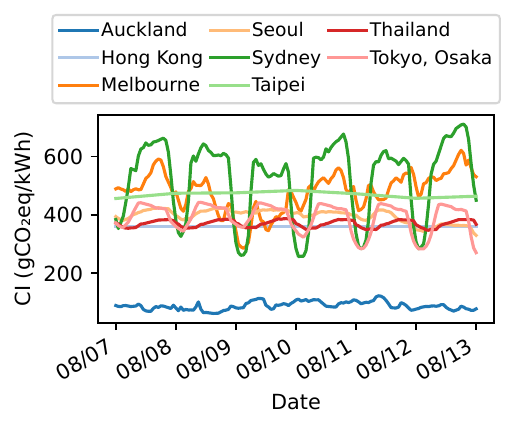}
        \vspace{-20pt}
        \caption{Asia-Pacific, Oceania}
        \label{fig:AP_CI_week}
    \end{subfigure}
\hfill
    \begin{subfigure}{0.49\columnwidth}
        \centering
        \includegraphics[width=\columnwidth]{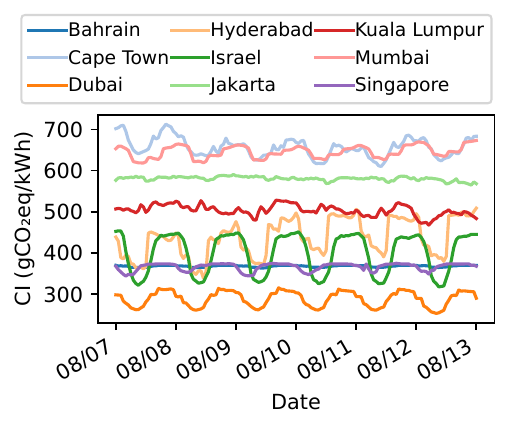}
        \vspace{-20pt}
        \caption{Africa, South(east) Asia}
        \label{fig:AF_CI_week}
    \end{subfigure}

    \vspace{-5pt}
    \caption{Hourly carbon intensity across AWS cloud regions. 
    }
    \vspace{-0.2in}
    \label{fig:CI_combined}
\end{figure}

\subsection{Microservice Applications}
\label{sec:back_ms}

Microservices represent a widely adopted architectural style in which applications are decomposed into small, self-contained services. Each service is responsible for a specific function, such as user authentication, data storage, or recommendation logic, and can be developed, deployed, and scaled independently. This modularity enables faster development cycles, fault isolation, and flexible scaling. 
Many real-world microservice applications, such as online banking, messaging platforms, e-commerce checkouts, and video streaming, are latency-sensitive, 
and must meet strict Service Level Objectives (SLOs)~\cite{DeathStar, uber-crisp,tao-fb}, such as P95 or P99 latency under 300~ms~\cite{parties}. 
This limits the extent to which services can be distributed across distant regions without degrading end-to-end latency.

\vspace{0.05in} 
\noindent{\bf Microservice Deployment Scale.} 
Large enterprises serving global user bases, such as Uber~\cite{microservices-uber2020_doma, microservices_use_cases}, Netflix~\cite{micro_techgiants, microservices_use_cases, emizentech_microservices_architecture}, and Meta~\cite{dcperf}, deploy {\it thousands} of microservices across the world. For example, Uber reports nearly 4{,}000 microservices~\cite{uber-crisp}. In contrast, one of the world's largest fintech companies, the Ant Group, deploys only 135 microservices~\cite{deepscaling}.
At the same time, enterprises that operate in geographically constrained deployments, typically have {\it tens to a few hundred} microservices. For example, the EU-based online bank N26~\cite{n26-1, n26-2} reports 180--280 services and the UK-based online grocery delivery Ocado~\cite{ocado} reports around 150 services. The Swedish IT consulting company Callista~\cite{callista} reports around 10 services~\cite{callista-ms, callista-git}. Other examples, such as Bokbasen in Norway~\cite{bokbasen, bokbasen-ms}, and Casavo from Italy~\cite{casavo, casavo-ms}, do not report their microservice count.
This difference in scale and geographic reach is an important factor in the design and evaluation of carbon-aware microservice deployments.

\subsection{Carbon Intensity}
\label{sec:background:carbon}

The environmental cost of computing is largely determined by the operational~\cite{sunk-carbon} carbon intensity of electricity in the region where workloads execute~\cite{caribou, Lechowicz2025, ecovisor}. Carbon intensity refers to the amount of carbon dioxide equivalent (CO$_2$eq) emitted per kilowatt-hour (kWh) of electricity produced. It varies significantly across geographical locations and time, depending on the local energy mix, i.e., the proportion of renewable versus fossil-fuel-based generation~\cite{electricity-map, ecofreq}. Figure~\ref{fig:CI_combined} shows hourly carbon intensity (CI) for major cloud regions across the world over one week in August 2023, using public data from Electricity Maps~\cite{electricity-map}. 

We observe that, across the world, clear day–night patterns appear: CI tends to drop around midday and rise in the evening due to demand cycles and renewable variability. For example, solar energy peaks at noon and declines after sunset, increasing reliance on fossil generation~\cite{electricity-map}. Regions with stable baseload power (e.g., nuclear, hydro) show minimal variation, while those with high renewable shares fluctuate more~\cite{electricity-map}.

Figure~\ref{fig:CI_combined} shows that some regions are consistently greener, such as Paris and Stockholm in Europe and Montreal in North America. However, this does not imply that a ``follow-the-sun'' strategy is optimal, as it assumes a global footprint capable of shifting workloads across far-apart regions~\cite{ecolife, green-for-less-green, lets-wait-awhile, caribou, carbonedge}. In practice, many organizations cannot move workloads across continents due to regulatory or infrastructure constraints.

\subsection{Financial Feasibility}

\begin{figure}[t]
    \centering
    \includegraphics[width=0.9\linewidth]{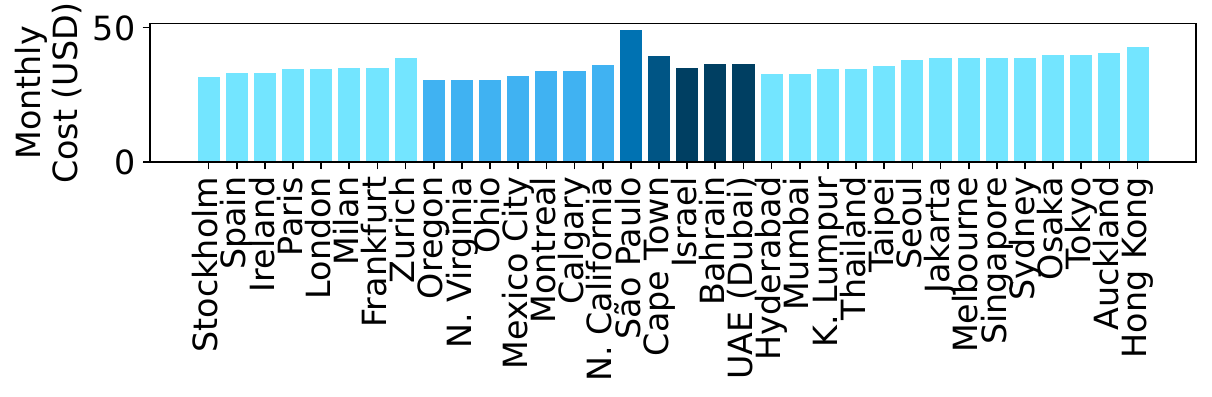}
    \vspace{-0.2in}
    \caption{Monthly cloud cost across global regions.
    }
    \label{fig:region_cost}
    \vspace{-0.2in}
\end{figure}

Figure~\ref{fig:region_cost} illustrates the monthly cost of an AWS \texttt{t3.medium} instance~\cite{aws-pricing} across global regions as a representative pricing trend. The narrow cost variation creates an important insight: {\it greener regions are not necessarily more expensive than carbon-heavy ones.} For example, Stockholm, which is one of the lowest-carbon regions, offers the lowest cost among EU options, making expansion to greener regions financially viable. 
Identifying a set of regions that enables carbon- and cost-efficient operation is crucial, particularly for SMEs, where even small price differences compound over many instances and long deployments. Expanding to greener regions without significant cost penalties allows SMEs to meet sustainability goals while maintaining financial efficiency.

\section{Insights from a Real Deployment}
\label{sec:motivation}

To investigate the feasibility of carbon-aware microservice deployment within geographically constrained infrastructures, we conduct a real-world experimental study. The goal of this study is to {\it extract insights} that will later inform the design of our adaptive microservice placement solution.

\begin{figure*}
    \centering
    
    \begin{subfigure}[t]{0.3\textwidth}
        \centering
        \includegraphics[width=0.99\linewidth]{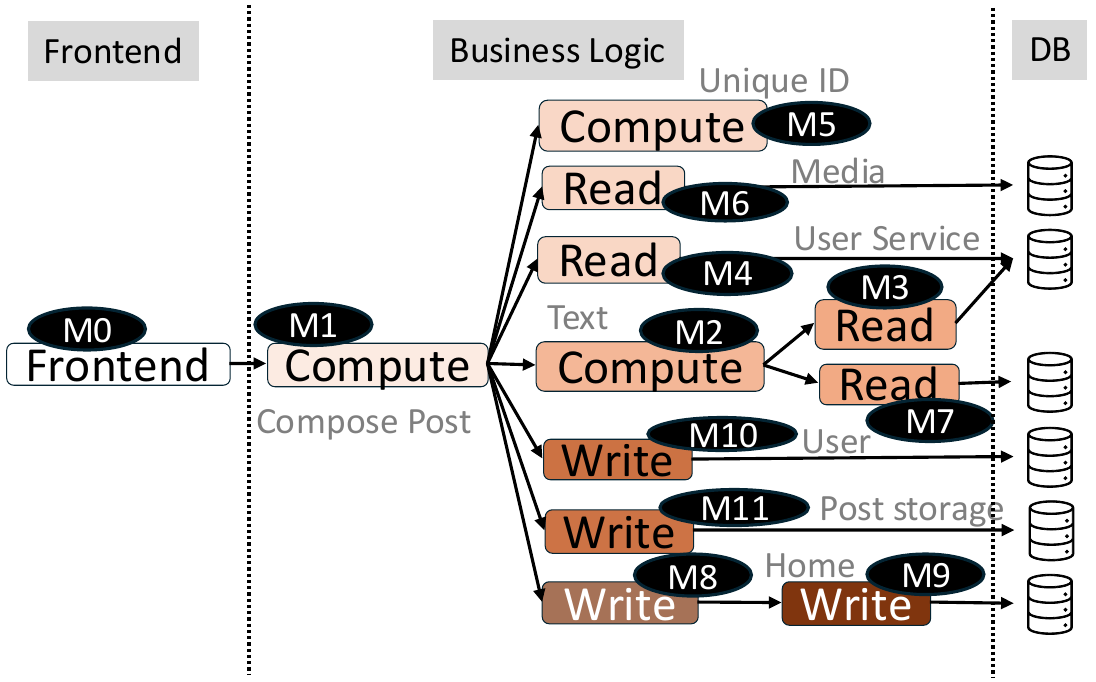}
        \vspace{-16pt}
        \caption{Social network.}
        \label{fig:MS_DAG}
    \end{subfigure}
    \hfill
    \begin{subfigure}[t]{0.4\textwidth}
        \centering
        \includegraphics[width=0.99\linewidth]{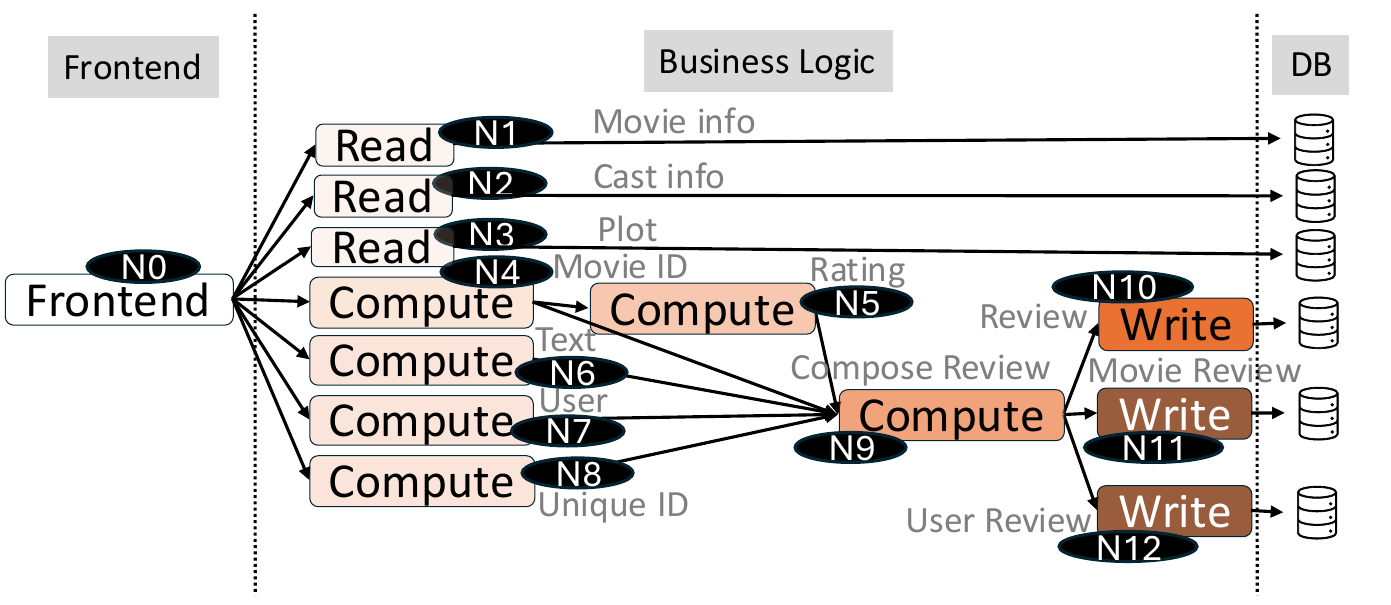}
        \vspace{-8pt}
        \caption{Media streaming.}
        \label{fig:MS_DAG_media}
    \end{subfigure}
    \hfill
    \begin{subfigure}[t]{0.29\textwidth}
        \centering
        \includegraphics[width=0.99\linewidth]{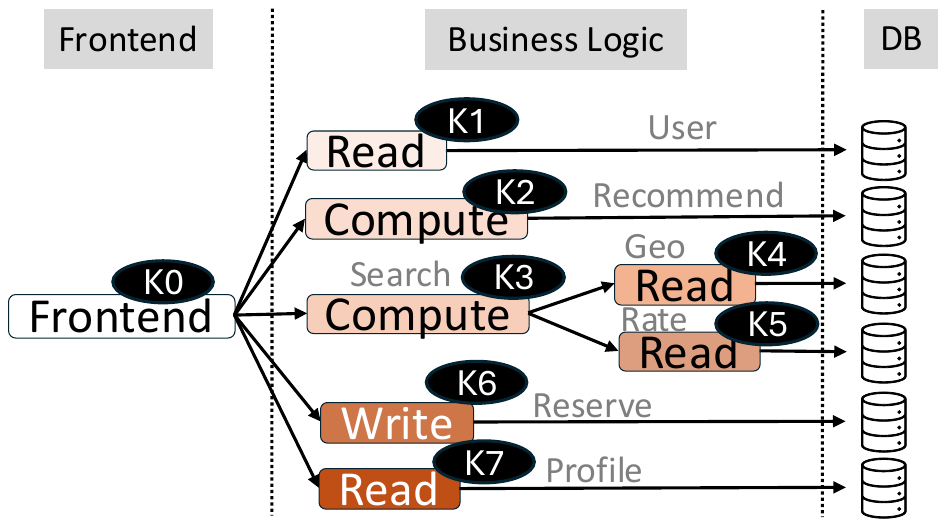}
        \vspace{-16pt}
        \caption{Hotel reservation.}
        \label{fig:MS_DAG_hotel}
    \end{subfigure}
    
    \caption{Architecture and functionality for three different applications from DeathStarBench~\cite{deathstar-link, DeathStar}.
    } 
    \label{fig:dags}
\end{figure*}

\begin{figure*}[h]
    \centering
\includegraphics[width=0.999\linewidth]{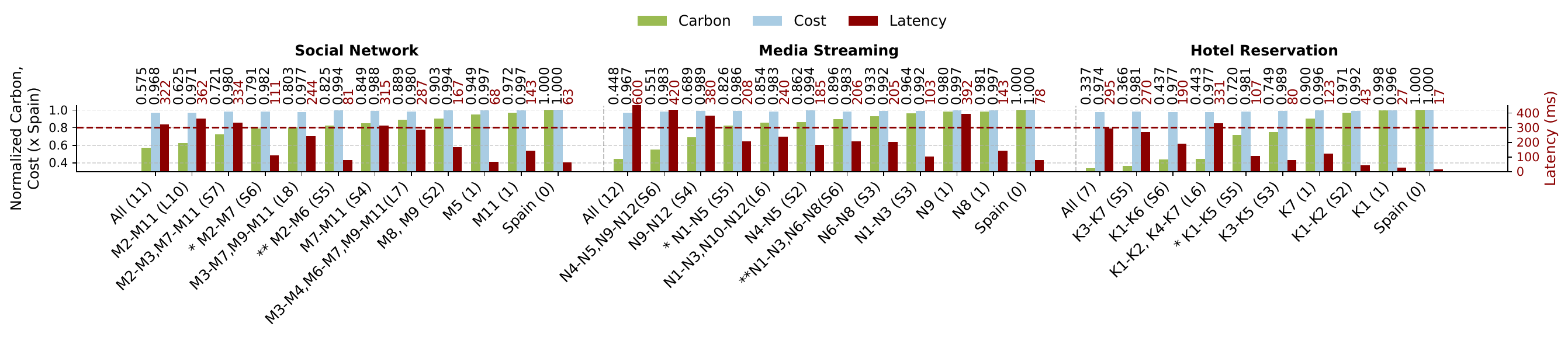}
    \vspace{-0.3in}
    \caption{Comparison of carbon,
    cost, and 
    latency across different placement configurations for the representative applications.
 }
    \label{fig:all_apps_motiv}
\end{figure*}

\vspace{0.05in}
\noindent{\bf Architecture of Microservice Applications.}
To enable a comprehensive motivational analysis, we deploy the DeathStarBench Social Network, Media Streaming, and Hotel Reservation applications~\cite{DeathStar, deathstar-link}. Figure~\ref{fig:dags} depicts the microservice dependency graphs of each application. 
Each microservice is identified by a node ID and colored according to its {\it activation order} in the request execution path, where darker shades indicate later stages and the same color denotes parallel execution.
Regarding application functionality, Social Network models composing and publishing posts, Media Streaming retrieves movie information and publishes user reviews, and Hotel Reservation supports searching and booking accommodations. All three follow the {\it same architecture} (Figure~\ref{fig:dags}): requests pass through the frontend, trigger business logic services to process data (e.g., user or item information), then application-specific services (e.g., composing posts, reviews, or reservations), and finally write results to backend databases. This modular design enables efficient deployment and scaling in cloud environments~\cite{imbres,ursa,nautilus, dagger, deeprest, DeathStar}. 

\vspace{0.05in}
\noindent{\bf Experimental Setup.}
We consider a European small to medium-sized enterprise (SME) headquartered in Spain, operating latency-sensitive microservice applications. \camready{The SME 
explores expanding its cloud deployment to greener regions but faces practical challenges such as tight budgets, geographically constrained operations, and user-facing applications with tight latency targets. Expanding to }
Sweden emerges as a promising opportunity due to its consistently low carbon intensity and competitive pricing. We focus on Spain and Sweden as illustrative regions: Spain represents a large, moderately carbon-intensive country with locally constrained deployment options, while Sweden offers a greener and more cost-effective alternative, as shown in Figures~\ref{fig:CI_combined} and~\ref{fig:region_cost}. This contrast allows us to isolate the impact of spatial offloading within a realistic EU context.
To assess regional placement trade-offs, we deploy the three applications on AWS \texttt{t3} instances in Spain and Sweden,
using the following static configurations: 
\begin{tightitemize}
    \item {\bf Single MS:} Move an individual 
    microservice at a time. 
    \item {\bf Subgraph:} Move a
    group of microservices,  
    taking into account the internal microservice dependencies. 
    \item {\bf Layer:} Move all microservices that are at the same layer, meaning at the same distance from the {\it databases} in our setup.
\end{tightitemize}


\vspace{0.05in}
\noindent{\bf Analysis of Results.} 
The x-axis labels in Figure~\ref{fig:all_apps_motiv} summarize static configurations across applications, ordered by increasing carbon reduction. Each label lists the microservices moved from Spain to Sweden, with parentheses indicating the number relocated, and \textit{S} and \textit{L} denoting subgraph and layer configurations (e.g., M2--M3, M7--M11 (S7) for a subgraph of seven services: M2, M3, M7, M8, M9, M10, M11). Asterisks (*) are explained later and in Section~\ref{sec:eval_baselines}. 
Carbon and cost are shown on the left y-axis normalized to the all-Spain baseline, while latency is on the right y-axis, with the horizontal line indicating the SLO. Lower bars indicate better performance. Due to latency and data sovereignty constraints, \textbf{frontend and databases} always remain in Spain.

Overall, we observe that normalized carbon and cost decrease as more services are moved to Sweden, with the largest reduction (43\%--66\% for carbon and $\approx$3\% for cost) achieved when relocating \textit{all} business logic. However, this comes at the cost of significantly higher latency, more than 2--3$\times$ that of the Spain-only baseline. In contrast, moving individual services yields only marginal carbon benefits. The best trade-off across all three metrics arises when relocating subgraphs, but the outcome depends mainly on 
the \textit{type} of services moved. For example, in Media Streaming, moving N4--N5 and N9--N12 yields lower carbon than moving N1--N3 and N6--N8, despite relocating the same number of services, as it shifts more compute- and write-intensive services to the greener region, but in return for much higher latency (420ms vs.\ 205ms).

Taking a deeper look into why this happens, a {\bf key insight} emerges: the \textbf{activation order} of the relocated microservices critically affects latency. Moving services later in the execution path introduces cross-region communication near the end of request processing, increasing latency, whereas relocating earlier services yields lower latency even with the same number of services. This behavior is {\it consistent across applications}: in Social Network (e.g., M2--M7 vs. M7--M11), Media Streaming (N1--N5 vs. N9--N12), and Hotel Reservation (K1--K5 vs. K3--K7).

The placements achieving the \textit{best trade-off} across carbon, cost, and latency are marked with a single * in Figure~\ref{fig:all_apps_motiv}: M2--M7 (S6) for Social Network, N1--N5 (S5) for Media Streaming, and K1--K5 (S5) for Hotel Reservation. They reduce carbon 
by 21\%, 17\%, and 28\%, while 
lowering monthly cost by 
1.8\%, 1.4\%, and 1.9\%. Latency remains well within the 300ms SLO~\cite{firm, imbres, wang2024autothrottle} at 111ms, 208ms, and 107ms, with SLO violations of just 1.67\%, 0.33\%, and 0.02\%. In contrast, configurations with lower normalized carbon exceed 26\% of SLO violations across all applications. Section~\ref{sec:evaluation} shows that these placements are also the ones selected by \sys.

\vspace{0.05in}
\noindent{\bf Summary.} 
Spatial offloading enables reductions in carbon and cost, but latency depends on {\it 
which services} are moved. In particular, offloading {\it early-activated} services allows for lower end-to-end latency than moving later ones, a behavior consistent across diverse microservice applications. \camready{This happens because moving an early-activated service can amortize its additional geographic latency over the remainder of the request execution.}  Overall, this complex interplay between carbon, cost and latency makes geo-distribution decisions inherently challenging, motivating our system design.

\section{\sys System Design}
\label{sec:system}

\subsection{Overview and Objectives}
\label{subsec:overview}

We introduce {\bf \sys}, a system that enables adaptive, carbon- and cost-aware microservice placement across geo-distributed regions for latency-sensitive applications. Unlike prior work that assumes access to global-scale infrastructure, 
\sys is designed for 
national or local companies and small to medium-sized enterprises (SMEs) operating within constrained geographic environments and policies. 
These organizations often face limited resource availability, cost budgets, strict data sovereignty regulations, and heterogeneous resource capacities, making traditional carbon-aware strategies impractical. \sys addresses these constraints directly, enabling sustainable microservice deployment without requiring global reach.

At its core, \sys solves a constrained optimization problem to determine where microservices should be placed across available regions. \sys balances three key {\bf design goals}:

\begin{itemize}[label={}, leftmargin=25pt, itemsep=0pt, topsep=0pt]
    \item[\textbf{[O1]}] \textbf{Carbon Efficiency:} Minimize emissions by leveraging regions with cleaner energy sources.
    \item[\textbf{[O2]}] \textbf{Cost Efficiency:} Lower the cost of expanding operations to greener regions.
    \item[\textbf{[O3]}] \textbf{Latency Compliance:} Ensure service-level objectives (SLOs) are consistently met.
\end{itemize}

\begin{figure}
    \centering
    \includegraphics[width=0.8\columnwidth]{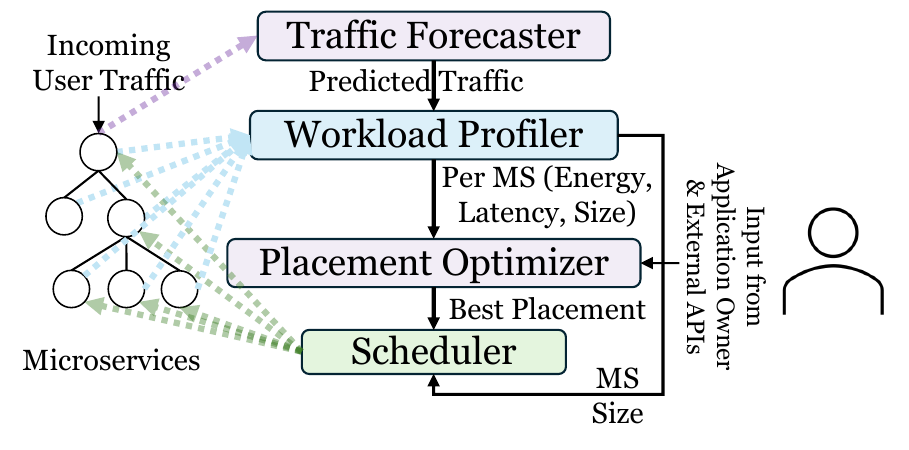}
    \vspace{-0.1in}
    \caption{System Components of \sys.}
    \vspace{-0.2in}
   \label{fig:system}
\end{figure}

\noindent Figure \ref{fig:system} illustrates the system components of \sys:

\begin{tightenumerate}
    \item {\bf Traffic Forecaster:} Uses a machine learning model to predict short-term traffic volumes from historical patterns (Section~\ref{subsec:predictor}).
    \item {\bf Workload Profiler:} Maps forecasted traffic to per-microservice metrics, such as latency, energy, and resource use, via hash-table lookups populated during offline profiling (Section~\ref{subsec:profiler}).
    \item {\bf Placement Optimizer:} Solves an 
    optimization problem using a scalable genetic algorithm, with search space pruning via region filtering and microservice pinning. It uses real-time carbon, cost, and inter-region latency data from external APIs (Section~\ref{subsec:optimizer}).
    \item {\bf Scheduler:} Enacts geo-distribution decisions by provisioning resources, powering on services from pre-replicated images, rerouting traffic once they are ready, and shutting down services in the original region. 
    (Section~\ref{subsec:scheduler}).
\end{tightenumerate}

\noindent

\vspace{0.05in}
\noindent{\bf Input Parameters.}
The following inputs are provided by the application owner to \sys:
\begin{tightitemize}
\item {\bf Application.} Access to the full microservice application, to manage placement and analyze inter-service dependencies.
\item {\bf Latency SLOs.} End-to-end application latency constraints.
\item {\bf Candidate regions.} A list of allowed deployment regions, constrained by regulatory or business requirements (e.g., GDPR).
\item {\bf Optimization weights.}
 Owners may specify the $w_{CO_2}$ and
$w_{\$}$ weights to prioritize carbon reduction, cost savings, or both.
The weights take fractional values between 0 and 1. If unspecified,
both weights default to 0.5, resulting in equal prioritization.
\end{tightitemize}

\vspace{0.05in}

\noindent{\bf Interfacing with the application owner.}
\sys exposes a terminal-based configuration interface. 
Owners provide inputs as key-value command-line arguments, enabling reconfiguration without code or file modifications. Updated parameters can be submitted at any time and \sys reloads and applies them immediately, avoiding system restarts. For each new placement decision, the owner receives a notification with associated carbon and cost metrics. Only structural application changes (e.g., new services or dependencies) require system reinitialization. This interface offers flexible and transparent control over cost, carbon, and performance trade-offs.

\vspace{0.05in}
\noindent{\bf System Initialization.} 
During initialization, \sys prepares each component for runtime operation. The Traffic Forecaster trains a short-term prediction model using historical request traces, while the Workload Profiler collects per-microservice performance metrics under representative workloads. This data collection and offline training phase spans 
one week to capture recurring traffic and usage patterns, which are predominant in cloud microservice workloads~\cite{alibaba21, imdea-midas, seagull, nautilus,imdea-is-ml-necessary, alibaba-socc22, cluster-workloads-atc}. 
The Placement Optimizer performs an initial region filtering step to exclude suboptimal deployment locations based on real-time carbon intensity and cost data. Meanwhile, the Scheduler pre-replicates container images across the filtered candidate regions to enable fast service migration. These steps are executed once before entering online mode.

\subsection{Traffic Forecaster}
\label{subsec:predictor}
The Traffic Forecaster predicts short-term user traffic to enable proactive and adaptive microservice placement, since prediction for short-term windows has been shown to be more accurate than longer-term forecasting~\cite{patchtst, nhits, sibyl}.
It 
continuously monitors ingress traffic at the entry point (e.g.,~nginx or Envoy Gateway), capturing traffic rates in real time.

\noindent
{\bf Offline Training.} During system initialization, \sys collects real user traces 
over several days. These are used to train one lightweight Gradient Boosted Decision Tree (GBDT) for short-term traffic prediction with a 1-hour forecasting horizon, matching the update frequency of regional carbon-intensity and pricing signals~\cite{electricity-map, aws-pricing}. As shown in Section~\ref{sec:eval_ml}, GBDTs are best suited for this setting because they handle tabular workload traces, capture non-linear patterns with little training data, and offer fast inference, making them a common choice in ML-for-systems workloads~\cite{lava-gbdt, kerveros, barista, statuscale}.
Forecasts are aggregated per interval, with the maximum used as the next-hour load estimate. This conservative approach filters transient bursts and prevents oscillations while remaining responsive to meaningful workload changes. Since latency-sensitive applications commonly exhibit 15–45 minute bursts~\cite{AzureTrace, ms-oscillation}, it improves stability without sacrificing adaptability.

\noindent
{\bf Online Inference.} At runtime, the forecaster predicts short-term traffic by running inference every hour. At timestamp $t$, it outputs $\text{Traffic}(t{+}1)$, representing the expected load for the next hour. Both the historical window and forecasting horizon use a 5-minute granularity, so each prediction is based on the most recent 12 samples (covering the past hour), as also done in~\cite{sibyl, dote-nsdi, shao2022pre, fan2020novel, roy2021sst, huo2023hierarchical}.

\noindent
{\bf Adaptivity.}
\sys performs periodic retraining during low-load periods (e.g., weekly at night) to maintain accuracy and adapt to gradual workload shifts, as often done in production~\cite{lava-gbdt, seagull, xu2025green}.


\subsection{Workload Profiler}
\label{subsec:profiler}

The Workload Profiler maps forecasted application traffic to per-microservice performance, resource, and energy estimates that are then used to guide placement decisions.

\noindent
{\bf Offline Profiling.} During system initialization, the profiler samples CPU, memory, network utilization, latency, and energy consumption for each microservice under the observed workload. These measurements populate a hash table indexed by traffic buckets, enabling fast lookups~\cite{imbres, ice-buckets}. Traffic measurements are aggregated over 5-minute intervals using average values.

To enable efficient lookup and generalization, \sys partitions the observed traffic range into up to $K=10$ equal-frequency buckets, with at least $N_{\text{min}}=100$ profiling samples per bucket. Specifically, $K=\min(10,\lfloor N/N_{\text{min}}\rfloor)$ with a minimum of $K_{\text{min}}=3$. Buckets that fail to meet the sample threshold are merged with adjacent buckets. 
For each bucket, the profiler computes a conservative representative value for each metric, that is the mean plus one standard deviation ($\mu + \sigma$), though upper percentiles (e.g., 85th percentile) may be used for stronger tail conservatism.  Each bucket stores $\mathcal{H}[\text{Traffic}(t), \text{MS}_i] = [E_i(t), L_i(t), S_i(t)]$ for each microservice $i$ and timestep $t$, where $E$ is energy, $L$ is internal latency, and $S = [\text{CPU}, \text{Mem}, \text{Net}]$ is the resource footprint. This stage of \sys reflects the workload rate in the final deployment, with resource allocations following profiler-derived demands to enable heterogeneous provisioning.

\noindent
{\bf Online Lookup.} At runtime, the profiler uses forecasted traffic to perform a millisecond-scale lookup in $\mathcal{H}(t+1)$, retrieving per-microservice metrics that enable the optimizer to evaluate placement feasibility, carbon impact, and cost.

\noindent{\bf Adaptivity.}
The Workload Profiler continuously refreshes its hash table, retaining one week of metrics and evicting the oldest entries once it exceeds 2000 samples, similar to~\cite{caribou}. This process is independent of retraining and software updates.


\subsection{Placement Optimizer}
\label{subsec:optimizer}

The Placement Optimizer assigns movable microservices to geo-distributed regions to minimize carbon emissions and cost while satisfying latency SLOs. It formulates a constrained optimization problem over a discrete set of microservices and candidate regions, minimizing a weighted sum of carbon and cost under latency and dependency constraints. 

Microservice placement is a hard combinatorial problem. It is NP-hard in general~\cite{ms-placement, ms-placement-joined-opt, iot-ms-placement}, but the bigger obstacle for \sys is non-linearity: latency depends on execution paths and inter-service dependencies, not on a linear function of the number of services~\cite{uber-crisp, graf, fastpert}. This makes exact linear solvers such as MILP a poor fit. In addition, \sys needs to be able to react quickly to dynamic triggers, which rules out exact solvers entirely, due to their prohibitive convergence times. Genetic algorithms~\cite{john2019adaptation, goldberg1989optimization}, by contrast, are known to handle well large-scale, constraint-heavy problems with non-linear objectives~\cite{cortellessa-ga, ga-app-placement}. \sys therefore adopts a GA-based approach.

The algorithm iteratively selects high-quality placements (those with low carbon, low cost and latency that satisfies the SLO), recombines them via crossover, and introduces diversity through mutation. Infeasible solutions are discarded, and the process continues until convergence, or a time budget is reached. This approach allows \sys to find near-optimal placements in a scalable way.

\vspace{0.05in}
\noindent{\bf Search space pruning.} \sys introduces {\it microservice pinning} and {\it region filtering} as key optimizations \camready{to reduce the search space and speed up the process of determining the region to microservice mappings. } 
\camready{These pruning mechanisms are particularly beneficial for geographically constrained SMEs, for which strict cost budgets and carbon objectives interact with data sovereignty and latency constraints, introducing additional complexity.} 

{\bf 1. Microservice Pinning} reduces the optimization search space by excluding microservices that must remain in their base region. First, \sys { \it  structurally pins} microservices that are inherently immovable due to operational or regulatory constraints, such as frontends that must remain close to users for responsiveness, and databases that are bound by data sovereignty laws~\cite{tao-fb,data-sovereignty-2024, data-survey-2024-peer-review}. \camready{Second, \sys leverages our insight from Section~\ref{sec:motivation}, that moving later-activated microservices 
introduces severe latency delays. 
Therefore, \sys ranks microservices based on their activation order and pins the ones that are later activated, marking a smaller percentage as ``movable''.}  
\camready{\sys begins pinning late-activated microservices when end-to-end latency in the base region approaches the SLO, for example when it reaches about 80–90\% of the SLO~\cite{parties}.}
Together, these mechanisms ensure that only latency-tolerant and legally movable microservices are considered for geo-distributed placement, effectively reducing the search space.

{\bf 2. Region Filtering} 
further reduces the search space by pruning suboptimal regions using real-time carbon intensity and pricing data. Region pruning follows a Pareto-style rule, which retains regions only if they do not worsen all metrics compared to the base region. To prevent cost increases, which is critical for SMEs, \sys applies a conservative dominance rule, when both carbon and cost are active objectives. 
If only carbon or only cost are active objectives, then only regions that improve on that metric are considered.  
Filtering occurs during initialization and before each optimization cycle, enabling \sys to adapt to evolving cost–carbon conditions.

\vspace{0.05in}
\noindent{\bf Optimization Execution.} For each candidate placement, that is the (microservice, region) combinations of the pruned search space, the optimizer calculates the following three metrics using inputs from the profiler, the application owner, and external APIs. First, it computes end-to-end latency by combining internal microservice latencies provided by the Workload Profiler with inter- and intra-region RTTs retrieved from external APIs~\cite{aws-latency}. Second, it computes total carbon emissions by using per-microservice energy usage from the Workload Profiler together with regional carbon intensity obtained from real-time carbon APIs. Those values are multiplied to derive the carbon footprint of each microservice, as proposed in related works~\cite{caribou, green}. Third, it computes deployment monetary cost by mapping resource demands supplied by the Workload Profiler to the smallest compatible cloud instance type, 
using live pricing information from the corresponding API. 
The {\it optimization result} is a region assignment $r$ for each microservice. In summary, the optimizer solves: 
\[
\begin{gathered}
\min_{r} \; w_{CO_2} \!\!\sum_{m \in \mathcal{M}} \!\text{CI}_{r(m)} E_m 
+ w_{\$} \!\left(\!\sum_{m \in \mathcal{M}} P_{r(m),s_m} + C_{\text{store}} + C_{\text{traffic}}\!\right)\\
\quad \text{s.t.} \; \text{Latency}(r) \le \text{SLO}
\end{gathered}
\] 

where $\mathcal{M}$ is the set of microservices and $r(m)$ denotes the region assigned to microservice $m$. $\text{CI}_{r(m)}$ is the carbon intensity of region $r(m)$ and $E_m$ the energy consumption of $m$. The cost term includes the compute cost $P_{r(m),s_m}$ for provisioning resources for the service $m$ with size $s_m$ in region $r(m)$, along with storage cost $C_{\text{store}}$ for maintaining pre-replicated container images across the filtered regions, and the inter-region data traffic cost $C_{\text{traffic}}$. 
The weights $w_{CO_2}$ and $w_{\$}$ are user-defined, and control the priority of the optimization objectives. The latency constraint guides the optimization toward SLO-compliant deployments.

The GA-based optimizer has time complexity $\mathcal{O}(M^2)$ and memory complexity $\mathcal{O}(R^2 + M^2)$, where $M = |\mathcal{M}|$ is the number of movable microservices and $R = |\mathcal{R}|$ is the number of candidate regions. Our insight-driven pruning reduces these to $\mathcal{O}(M'^2)$ and $\mathcal{O}(R'^2 + M'^2)$ with $M' < M$ and $R' < R$, leaving the asymptotic Big-O unchanged while lowering the constants. For example, if microservice pinning reduces $M$ by one-third ($M' = \tfrac{2M}{3}$) and region filtering halves $R$ ($R' = \tfrac{R}{2}$), runtime becomes $(\tfrac{2}{3})^2 = \tfrac{4}{9}$ of the original (approximately $44.4\%$). Memory becomes $\tfrac{1}{4}R^2 + \tfrac{4}{9}M^2$, i.e., between $\tfrac{1}{4}$ and $\tfrac{4}{9}$ of the original depending on whether region or microservice terms dominate.

\vspace{0.05in}
\noindent{\bf Event-driven Optimization.}
While the Traffic Forecaster and Workload Profiler run periodically, the Placement Optimizer is triggered by carbon or workload changes.
   
\textit{Carbon-driven trigger:} \sys monitors carbon intensity forecasts via CarbonCast~\cite{carboncast}, a widely used solution for predicting carbon intensity~\cite{ecovisor, green-for-less-green, carbon-scaler, casper}. \sys adopts a pluggable design that allows integrating alternative or more recent predictors (e.g., LiteCast~\cite{LiteCast}) as they become available. 
Optimization is triggered when a region’s predicted carbon intensity deviates significantly (more than 20\%) from its recent trend~\cite{xu2025green, bostandoost2024data, maji2023bringing}, as it happens due to a transition from day to night and vice versa (Figure~\ref{fig:CI_combined}). 

\textit{Workload-driven trigger:} The optimizer reacts to forecasted traffic changes. When predicted load shifts to a new bucket, indicating a significant deviation from current capacity, it triggers reoptimization. 
For example, if the system is deployed for medium traffic (100–500 requests/sec) at 300 requests/sec, and forecasts an increase to high traffic (500-700 requests/sec) with 600 requests/sec, this shift triggers reoptimization to adjust to the anticipated higher load.

\subsection{Scheduler}
\label{subsec:scheduler}

The Scheduler enacts the placement plan produced by the Optimizer. For each $(\text{microservice}, \text{region})$ assignment, the scheduler provisions resources in the target region and instantiates the corresponding microservices from pre-replicated images. The images were pre-replicated during \sys's initialization to enable fast transitions in geo-distributed placement at minimal cost (e.g., AWS ECR storage costs \$0.10/GB per month~\cite{aws-ecr}). The scheduler next verifies microservice readiness through control-plane health and dependency checks before rerouting traffic to the new instances, ensuring SLO compliance. Once stable, the scheduler decommissions old instances and performs resource cleanup.

\begin{figure*}[t]  
    \centering

    \begin{subfigure}[t]{\linewidth}
        \centering
        \small
        \begin{tabular}{p{2cm}|c|p{9cm}|p{3cm}}
            \toprule
            {\bf Base Location} & {\bf Initial Regions} & {\bf Filtered Regions:} $\leq$ carbon intensity (gCO2e/kWh), compute cost (\$) than the base location.  & {\bf Used Regions:} $\leq CO_2$, {\it total} \$ {\it and} latency constraint \\ 
            \midrule
            {\bf EU:} Frankfurt & All in the world & {\bf N. America:} Montreal. {\bf EU:} London, Paris, Spain, Stockholm.  & {\bf EU:} Paris, Stockholm \\ \hline
            \makecell[l]{{\bf N. America}: \\Calgary} & All in the world &   
            \makecell[l]{{\bf N. America:} Mexico, Montreal, N. Virginia, Ohio, Oregon.\\
            {\bf EU:} Ireland, Spain, Stockholm. } & {\bf N. America:} Montreal \\ \hline
            {\bf Asia}: Hong Kong & All in the world & \makecell[l]{{\bf N. America:} Calgary, Mexico, Montreal, N. Californ., N. Virginia, Ohio, Oregon. \\
            {\bf EU:} Frankfurt, Ireland, London, Milan, Paris, Spain, Stockholm, Zurich.  \\
            {\bf Asia:} Bahrain, Hyderabad, Israel, Kuala Lumpur, Mumbai, Osaka, Seoul, Taipei, \\
            Thailand, Tokyo, UAE. {\bf Oceania:} Auckland, Melbourne.} &  {\bf Asia:} Kuala Lumpur, Seoul. {\bf Oceania:} Auckland \\
            \bottomrule
        \end{tabular}
        \caption{\sys's Region Selection Process: Initial $\rightarrow$ Filtered $\rightarrow$ Used Regions for each scenario.}
        \label{tab:region-selection}
    \end{subfigure}
    
    \vspace{0.1in}
    
    \begin{subfigure}[t]{0.33\linewidth}
        \centering
        \includegraphics[width=\linewidth]{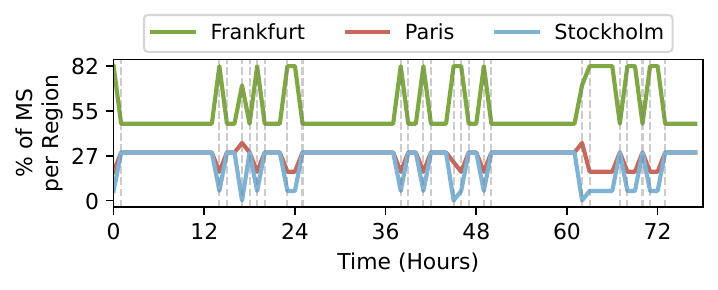}
        \vspace{-20pt}
        \caption{Europe - Social Network.}
        \label{fig:real-ms-timeseries-eu}
    \end{subfigure}
    \hfill
    \begin{subfigure}[t]{0.33\linewidth}
    \centering
        \includegraphics[width=\linewidth]{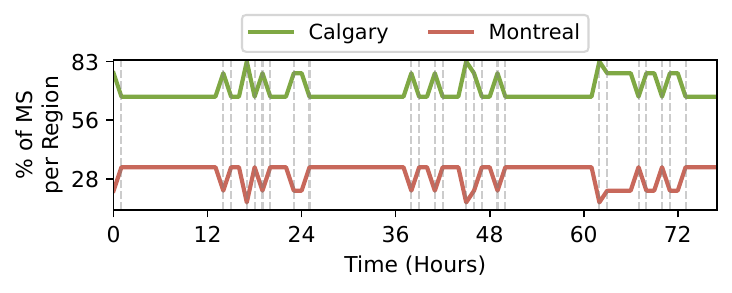}
        \vspace{-20pt}
        \caption{North America - Social Network.}
        \label{fig:real-ms-timeseries-us}
    \end{subfigure}
    \hfill
    \begin{subfigure}[t]{0.32\linewidth}
    \centering
        \includegraphics[width=\linewidth]{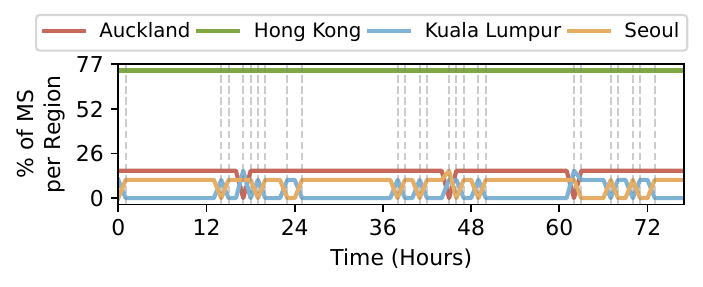}
        \vspace{-20pt}
        \caption{Asia - Media Streaming.}
        \label{fig:real-ms-timeseries-asia}
    \end{subfigure}

    \vspace{-0.1in}
    
    \caption{Microservice geo-distribution over time and breakdown of \sys's region selection per base location. 
    }
    \label{fig:combined-regions-real}
\end{figure*}

\subsection{ Implementation Details} 
\label{subsec:implementation}

\camready{\sys integrates AWS cloud-native tools and APIs across its components. AWS is used as an example of a public cloud provider already rented by an SME, not owned infrastructure. \sys can be instantiated on any cloud provider. SMEs commonly use public clouds to offload operational management (e.g., the German SME Lexoffice~\cite{lexoffice}). }
\begin{tightitemize}
    \item \textbf{Traffic Forecaster:}  
    Runs as a sidecar container next to the ingress gateway (NGINX or Envoy). It uses \texttt{Prometheus}~\cite{prometheus} to collect traffic metrics and runs a lightweight ML model deployed with scikit-learn~\cite{pedregosa2011scikit} and NeuralForecast~\cite{garza2022neuralforecast}, a PyTorch-based library for advanced time series forecasting.

    \item \textbf{Workload Profiler:}  
    Runs as a centralized database pod that stores microservice-specific telemetry. It samples CPU, memory, network, latency, and energy metrics via \texttt{Prometheus}, using \texttt{Node Exporter} for resource metrics and \texttt{Kepler} for energy data, and organizes them in a hash table indexed by traffic buckets.

    \item \textbf{Placement Optimizer:}  
    Runs as a control-plane pod executing a custom Python-based Genetic Algorithm. It queries the Electricity Maps API~\cite{api-electricity-map} for carbon intensity, the AWS Pricing Calculator~\cite{aws-pricing} for compute costs, and the AWS Latency Monitoring API~\cite{aws-latency} for inter-region RTTs. 

    \item \textbf{Scheduler:}  
    Runs as a control-plane pod. It uses Kubernetes for 
    scheduling, Liqo~\cite{liqo} for multi-region orchestration, and the Vertical Pod Autoscaler~\cite{vpa} for autoscaling. Microservices are deployed via Amazon Elastic Kubernetes Service
    ~\cite{aws-eks} on Elastic Compute Cloud (EC2) instances, with container images distributed through Amazon Elastic Container Registry (ECR)~\cite{aws-ecr}. 
\end{tightitemize}

\noindent

\section{Evaluation}
\label{sec:evaluation}

We evaluate \sys to demonstrate its ability to meet the three system design goals introduced in Section~\ref{sec:system}: (\textbf{O1}) carbon efficiency, (\textbf{O2}) cost efficiency, and (\textbf{O3}) latency compliance. 
Our evaluation covers (i) real-world deployment, (ii) comparison against state-of-the-art baselines, (iii) scalability, (iv) ablation of key design components,(v) sensitivity analysis, and (vi) geographic generalizability, providing a comprehensive assessment.

\subsection{Real-world Deployment}
\label{sec:eval_real}

\camready{We evaluated \sys using} the Social Network and Media Streaming applications from DeathStarBench~\cite{DeathStar} \camready{deployed on Kubernetes clusters across various AWS regions.} DeathStarBench is the standardized and most widely adopted benchmark for evaluating systems for microservice applications~\cite{cortellessa-ga, imbres, firm, wang2024autothrottle, cilantro, dagger, sinan, aquatope, erms, sage}. The experiment spans four consecutive days to evaluate \sys under realistic operational variability. Incoming requests are generated according to a Poisson process, with arrival rates scaled based on transition patterns observed in Azure Functions production traces~\cite{AzureTrace}, a trace commonly used to evaluate related works~\cite{ecolife,caribou,scaling-batch,microservices-azure,usher-azure,m-serve-azure}. 

\vspace{0.05in}
\noindent
{\bf Diverse Region Selection.} 
We consider three deployment scenarios with base locations (i.e., sources of user traffic) across {\it different continents}, optimizing for both carbon and cost. Specifically, we use Frankfurt (Europe) and Calgary (North America) for the Social Network application, and Hong Kong (Asia) for the Media Streaming application. Table~\ref{tab:region-selection} summarizes \sys's region selection process, showing how \sys starts from all cloud regions, filters out non-optimal ones {\it effectively reducing the search space}, and selects the geographically nearby regions that enable latency-aware (SLO = 300ms~\cite{parties, imbres}) carbon and cost reductions.

\vspace{0.05in}
\noindent{\bf Latency-aware Geo-distribution.}
Figure~\ref{fig:combined-regions-real} depicts the distribution of microservices across regions and marks with vertical dashed lines the times where the Placement Optimizer was triggered due to load fluctuations. In the Social Network deployments based in EU and North America, during predicted load spikes, most services return to their base locations to preserve latency compliance. 
When load stabilizes, \sys resumes offloading to greener regions such as Stockholm, Paris and Montreal. 
On average, \sys triggers 6--7 placement changes per day, relocating approximately 42\% of microservices per change in the European setting and 12\% in the North America setting. 
In contrast, in the Media Streaming deployment in Asia, the same 77\% of microservices need to remain in Hong Kong to maintain low end-to-end latency, even under varying load. A small subset of early-activated services is offloaded across other regions, as they can reduce carbon and cost while incurring acceptable latency overhead. 
Regarding latency, in Europe \sys achieves for Social Network an average end-to-end latency of 94ms and only 0.2\% SLO violations, in North America 40ms latency and 0.1\% violations, and in Asia, for Media Streaming, an 143ms average latency and 0.26\% SLO violations. The reported latencies are end-to-end and include data transfer between microservices.

\begin{figure}[t]
    \centering
    \begin{subfigure}[t]{0.49\linewidth}
        \centering
        \includegraphics[width=\linewidth]{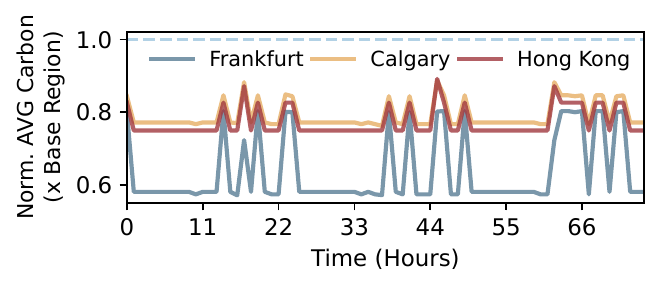}
        \vspace{-20pt}
        \caption{Carbon.} 
        \label{fig:real-carbon}
    \end{subfigure}%
    \hfill
    \begin{subfigure}[t]{0.49\linewidth}
        \centering
        \includegraphics[width=\linewidth]{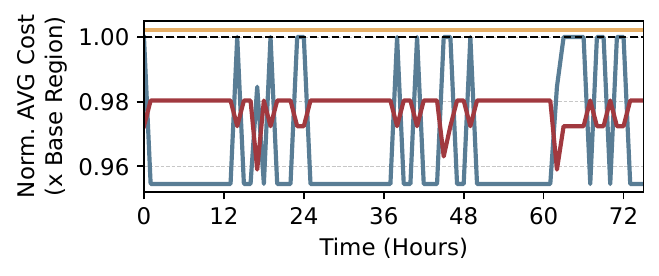}
        \vspace{-20pt}
        \caption{Cost.}
        \label{fig:real-cost}
    \end{subfigure}
    \centering
    \vspace{-0.1in}
    \caption{Carbon and cost reduction relative to the base-region-only deployment across the three scenarios.}
    \label{fig:real-aceso}
    \vspace{-0.1in}
\end{figure}

\vspace{0.05in}
\noindent{\bf Carbon Reduction.}
Figure~\ref{fig:real-carbon} presents the normalized carbon emissions over a 72-hour period for \sys compared to the static baseline that keeps all microservices in their base location. Across time, \sys achieves an average carbon reduction of 37.4\% in EU (Frankfurt), 21.2\% in North America (Calgary) and 23.2\% in Asia (Hong Kong). 
The higher carbon savings in EU are due to the availability of nearby low-carbon regions (e.g., Stockholm and Paris), enabling more microservices to be offloaded within the latency SLO. In North America, savings are lower, as fewer nearby regions offer similarly favorable carbon–latency trade-offs near Calgary. In Asia, despite having fewer geo-distributed services in different regions, 
\sys still achieves notable carbon reductions since Hong Kong has significantly higher carbon intensity than the other regions.

\begin{table}[t]
\centering
\small
\begin{tabular}{l|c|c|c|c}
\toprule
& \multicolumn{4}{c}{\textbf{Baseline / Aceso}} \\
\textbf{Deployment} & \textbf{Compute} & \textbf{Storage} & \textbf{Traffic} & $\Delta$\textbf{Cost (\%)} \\
\midrule
Frankfurt & 201.6\$ / 194.4\$ & -- / 0.19\$ & -- / 0.24\$ & -3.36\% \\
Calgary   & 187.2\$ / 187.2\$ & -- / 0.29\$ & -- / 0.12\$ & +0.22\% \\
Hong Kong & 439.2\$ / 424.8\$ & -- / 1.78\$ & -- / 0.18\$ & -2.83\% \\
\bottomrule
\end{tabular}
\caption{Monthly average cost breakdown of Baseline (all services in the base region) vs. \sys across deployments: compute (instance provisioning), storage (pre-replicated images), and cross-region traffic (request execution).
}
\vspace{-0.3in}
\label{tab:cost-breakdown}
\end{table}

\vspace{0.05in}
\noindent{\bf Cost Breakdown and Reduction.}
\sys does not migrate running microservices; instead, it instantiates them from pre-replicated container images and redirects traffic once they are ready. In Deathstarbench applications, any state is externalized to the {\it stateful} databases~\cite{DeathStar, dagger, ursa, firm}. This is common practice in microservice architectures~\cite{medium-stateless, bytebytego-stateless, digital-alpha-stateless, medium-stateless-guide, medium-stateless-success, medium-stateless-design}, although some applications may retain state within services~\cite{automq-stateless-vs-stateful, devopedia-stateless-vs-stateful, dremio-stateless-vs-stateful}, which is not the case here. \sys pins databases to the base location by design. Consequently, data transfer arises only from cross-region communication between {\it stateless} services along the request execution path.

Table~\ref{tab:cost-breakdown} reports the {\it monthly} cost breakdown across deployments. 
Compute cost sums up the cost of having one instance~\cite{aws-pricing} per service, sized to match the peak workload in each region during the time between workload-driven triggers. 
Storage incurs a small fixed monthly cost, because the size of all pre-replicated images remains below the minimum allocatable storage (1\,GB)~\cite{docker}.  
Traffic cost reflects inter-region communication based on data volume and per-GB pricing~\cite{aws-data-transfer}, while intra-region traffic is free. Absolute costs reflect the scale of the evaluated applications (Figure~\ref{fig:dags}).

Overall, \sys reduces monthly cost by 3.36\% and 2.83\% in Frankfurt and Hong Kong, respectively. The slight cost increase in Calgary (0.22\%) reflects \sys's {\it joint} optimization objective: the Placement Optimizer accepts a marginal cost increase (due to storage and inter-region communication) when it enables large carbon reductions.
Figure~\ref{fig:real-cost} shows the same cost reduction at a finer granularity, illustrating its variation over three days.

\subsection{Comparison Against Baselines}
\label{sec:eval_baselines}

We compare \sys against a diverse set of baselines that represent common strategies for microservice placement:

\begin{tightitemize}
    \item \textbf{Nautilus~\cite{nautilus}}:  dynamically reallocates microservices across nodes within the same geographical region to preserve QoS. {For our purposes,} this serves as a single region static  
    baseline without geo-distributed placement, and represents latency-aware microservice placement and load balancing techniques, such as Oakestra~\cite{oakestra}, Imbres~\cite{imbres}, FIRM~\cite{firm}. 
    
    \item \textbf{Caribou~\cite{caribou}}: applies heuristic-biased stochastic sampling for carbon-aware spatial shifting of serverless workloads. It considers carbon and cost but lacks fine-grained latency control required for microservices. Caribou represents SOTA solutions for carbon- and cost-aware scheduling. 
    
    \item \textbf{Linear Programming (LP)}: We include LP as an oracle baseline to approximate optimal placement under carbon, cost, and latency constraints. The problem is inherently non-linear, as described in Section~\ref{sec:motivation}, making exact LP formulations intractable at scale. Following standard relaxation techniques~\cite{feng2017approximation, yang2019delay, baev2008approximation}, we construct a linearized LP approximation.
    
    \item \textbf{Vanilla Genetic Algorithm (GA)}: A baseline inspired by prior GA-based optimizers~\cite{cortellessa-ga}, applying standard evolutionary search over the full placement space without any pruning.
    
\end{tightitemize}

\vspace{0.05in}
\noindent{\bf Real-world experimental validation.} 
We revisit the motivational experiment of Section~\ref{sec:motivation}, where a Social Network, a Media Streaming, and a Hotel Reservation application from DeathStarBench~\cite{deathstar-link,DeathStar} were deployed across Spain (base region) and Sweden. Placements by \sys, GA, and LP are marked with a single *, while Caribou with **.
For the initial static placement, the bottom of Figure~\ref{fig:all_apps_motiv} shows that \sys converges to the same static placement as the LP-based oracle and vanilla GA, confirming its ability to find near-optimal solutions under real-world constraints. Caribou produces similar configurations for the Social Network and Media Streaming and the exact same configuration for the Hotel Reservation app (marked with a single *). Figure~\ref{fig:all_apps_motiv} illustrates that \sys always finds the placement that yields the largest carbon benefit with very low SLO violations rates.
To extend this validation beyond static placements, we next rely on larger-scale, simulation-based experiments for dynamic scenarios.

\vspace{0.05in}
\noindent{\bf Simulation Environment.} 
We develop a simulation environment that models multi-region cloud infrastructures and large-scale microservice applications to enable scalable experimental analysis and comparison against other baselines. Simulation-based evaluation is commonly used in prior work on microservices and carbon-aware systems~\cite{carbonedge, lets-wait-awhile, carbon-scaler, ecolife, green-for-less-green, oakestra, imbres, wolski2023casper} (Table~\ref{tab:related_works}).  

As these systems do not model cross-region MS deployment, or do not publicly release their simulators, we base our design on the properties and functionality described in prior work on microservice call graph modeling~\cite{alibaba21, call-graphs-ieee}, as well as established benchmarks such as DeathStarBench~\cite{deathstar-link, DeathStar}.

The simulator mirrors the configuration and functionality of \sys. It takes the same inputs, including base and candidate regions, latency SLO, and optionally optimization weights. Instead of a concrete application, it models a generic microservice application specified by its size (number of services). It then simulates all \sys components under a workload-driven trigger, provided as a request rate change (requests/sec). The simulator outputs the resulting geo-distribution of services, along with aggregated carbon, cost reduction, and resulting latency.

Under the covers, the simulator models both sequential and parallel execution paths, enabling concurrent execution of independent services while preserving sequential dependencies, and captures per-microservice power consumption, load and latency based on the type of service e.g., frontend, business logic, database. It incorporates real-world latency measurements across AWS regions~\cite{aws-latency}, 
carbon intensity~\cite{electricity-map}, and pricing data~\cite{aws-pricing}, as well as empirical execution profiles derived from real measurements of representative microservice types using DeathStarBench~\cite{deathstar-link}, capturing how their latency and power consumption scales with different levels of load.

We validate the simulator's {\bf fidelity} by reproducing all above real-world experiments,
observing an average error of 1\% and 1.17\% for carbon, 0.46\% and 2.1\% for cost, and a deviation of 35ms and 41ms in average latency, respectively. The execution time of the simulation ranges from subseconds up to a few minutes, since it is dominated by the solve time of the optimization algorithm, which scales with the number of microservices. The simulator includes implementations of the evaluated baselines and is extensible, enabling the integration of additional optimization techniques.

\vspace{0.05in}
\noindent{\bf Simulation-based experiments.} 
We simulate an EU-based microservice application with 100 services, 
using the same workload as in the real-world experiment. Frankfurt is the base location, representing medium carbon intensity (Figure~\ref{fig:CI_combined}) and medium–high cost (Figure~\ref{fig:region_cost}). 
Only for this experiment, we relax the latency SLO, to allow the optimization of all baselines to converge.

\begin{figure}[t]
    \centering
    \begin{subfigure}[t]{0.48\linewidth}
        \centering
        \includegraphics[width=\linewidth]{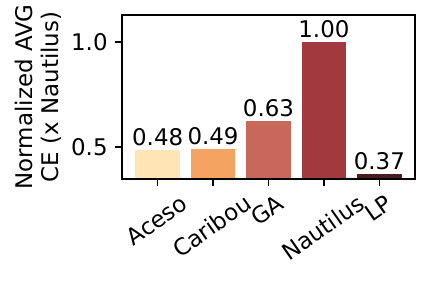}
        \vspace{-20pt}
        \caption{Carbon Emissions.}
        \label{fig:carbon-baselines}
    \end{subfigure}%
    \begin{subfigure}[t]{0.48\linewidth}
        \centering
        \includegraphics[width=\linewidth]{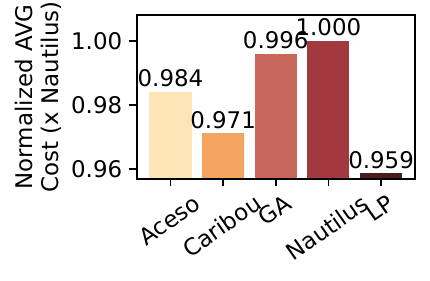}
        \vspace{-20pt}
        \caption{Monthly Cost.}
        \label{fig:cost-baselines}
    \end{subfigure}
    
    \begin{subfigure}[t]{0.48\linewidth}
        \centering
        \includegraphics[width=\linewidth]{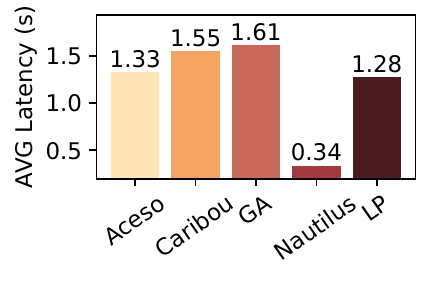}
        \vspace{-20pt}
        \caption{End-to-End Latency.}
        \label{fig:latency-baselines}
    \end{subfigure}
        \begin{subfigure}[t]{0.48\linewidth}
        \centering
        \includegraphics[width=\linewidth]{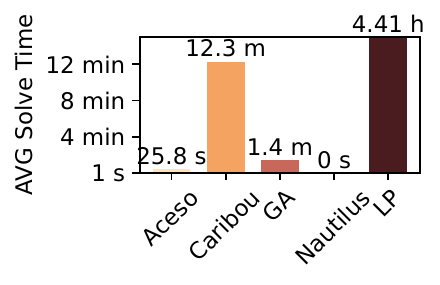}
        \vspace{-20pt}
        \caption{Solve Time.}
        \label{fig:time-baselines}
    \end{subfigure}%
    \centering
    \vspace{-5pt}
    \caption{Comparison against baselines.
    }
    \label{fig:baselines-comparison}
    \vspace{-0.2in}
\end{figure}

\vspace{0.05in}
\noindent\emph{Carbon Reduction.}
Figure~\ref{fig:carbon-baselines} shows the normalized average carbon emissions, relative to Nautilus, which keeps all microservices in Frankfurt. \sys achieves the best practical carbon savings, reducing emissions to 0.48$\times$ Nautilus, slightly outperforming Caribou (0.49$\times$) and significantly improving over vanilla GA (0.63$\times$). LP obtains the lowest emissions overall (0.37$\times$). 

\vspace{0.05in}
\noindent\emph{Cost savings.}
Figure~\ref{fig:cost-baselines} reports normalized average cost relative to Nautilus, accounting for compute resources, storage, and data transfer. 
All methods show similar cost savings (within 4\%). \sys achieves 0.984$\times$ Nautilus, slightly higher than Caribou (0.971$\times$) and LP (0.959$\times$), but better than vanilla GA (0.996$\times$).

\vspace{0.05in}
\noindent\emph{End-to-end latency.}
Figure~\ref{fig:latency-baselines} compares the average end-to-end latency across methods. For an application with 100 microservices, \sys achieves 1.33\,s, outperforming Caribou (1.55\,s) and GA (1.61\,s), while LP obtains slightly lower latency at 1.28\,s. Nautilus shows the lowest latency (0.34\,s), as it does not geo-distribute services.

\vspace{0.05in}
\noindent\emph{Solve time.}
Figure~\ref{fig:time-baselines} reports the average optimization solve time for each method. \sys converges in 25.8\,s, significantly faster than Caribou (12.3\,min) and GA (1.4\,min), while LP is impractical at 4.41\,h. We also tested whether \sys’s search space pruning strategy would accelerate LP convergence time. With pruning, LP converges in more than 30 minutes, which is still significantly slower than the other baselines. This highlights that it is the genetic algorithm that enables fast convergence, while the pruning strategy further accelerates and improves the solution of the optimization. 
Nautilus incurs zero solve time as it does not perform optimization.

\vspace{0.05in}
\noindent\emph{Geo-distribution.}
Table~\ref{tab:avg-ms} shows the average distribution of microservices per region. Nautilus keeps all services in Frankfurt (100\%), while LP aggressively offloads to low-carbon regions, mainly Stockholm (65.4\%) and Paris (9.6\%). Caribou and vanilla GA spread services across the most regions. In contrast, \sys prunes inefficient options, like Ireland and Milan, and prioritizes greener regions such as Paris (39.9\%) and Stockholm (23.9\%), achieving a more balanced placement that reduces both carbon and cost. 

\vspace{0.05in}
\noindent\emph{Responsiveness and stability.}
Table~\ref{tab:avg-response} summarizes responsiveness and stability across baselines. Over one day, 41 optimization triggers occur. \emph{Adaptation Rate} captures how often a new placement is produced: \sys, GA, and LP react to all triggers, while Caribou responds to only 16.6\% due to more conservative placement decisions during low load. \emph{MS moved} measures the fraction of relocated services. Caribou moves all movable services, risking instability, whereas LP moves only 3\%. \sys strikes a balance, moving 52.2\% on average (vs.\ 68.3\% for GA), thus avoiding excessive changes while remaining responsive to workload and carbon fluctuations.

\begin{table}[t]
\centering
\small
\begin{subtable}[t]{0.95\linewidth}
\centering
\begin{tabular}{|l||c|c|c|c|c|}
\hline
{\bf Region} & {\bf Aceso} & {\bf Caribou} & {\bf GA} & {\bf Nautilus} & {\bf LP} \\\hline
Frankfurt & 24.4\% & 25.6\% & 32.1\% & 100.0\% & 25.0\% \\\hline
Paris & 39.9\% & 17.0\% & 26.3\% & - & 9.6\% \\\hline
Stockholm & 23.9\% & 45.0\% & 20.2\% & - & 65.4\% \\\hline
London & 9.0\% & 2.0\% & 4.0\% & - & - \\\hline
Spain & 2.8\% & 1.0\% & 4.0\% & - & - \\\hline
Ireland & - & 2.0\% & 1.8\% & - & - \\\hline
Milan & - & - & 1.4\% & - & - \\\hline
\end{tabular}
\caption{Average percentage of microservices per region across time.}
\label{tab:avg-ms}
\end{subtable}

\vspace{0.05in}

\begin{subtable}[t]{0.95\linewidth}
\centering
\begin{tabular}{|l||c|c|c|c|c|}
\hline
{\bf Metric} & {\bf Aceso} & {\bf Caribou} & {\bf GA} & {\bf Nautilus} & {\bf LP} \\\hline
Adaptation Rate & 100\% & 16.6\% & 100\% & - & 100\% \\\hline
AVG MS moved & 52.2\% & 80\% & 68.3\% & - & 3\% \\\hline
\end{tabular}
\caption{Responsiveness and stability metrics.}
\label{tab:avg-response}
\end{subtable}

\caption{Microservice placement across regions and baselines.}
\vspace{-0.2in}

\end{table}

\vspace{0.05in}

\takeaways{}
\sys achieves carbon and cost savings comparable to the optimal Linear Programming (LP) baseline, while converging in seconds, orders of magnitude faster than other baselines. This is possible because \sys leverages a Genetic Algorithm (GA) to efficiently explore the large search space, while its effective pruning strategy focuses the search on high-impact regions and microservices, thus achieving better results than vanilla GA.

\begin{figure}[t]
    \centering
    \begin{subfigure}[t]{0.47\linewidth}
        \centering
        \includegraphics[width=\linewidth]{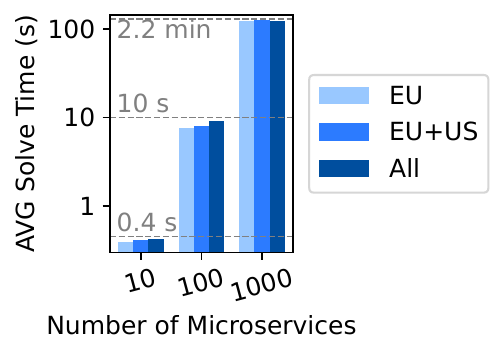}
        \vspace{-17pt}
        \caption{Solve time vs. microservice count and region set.}
        \label{fig:time-scale}
    \end{subfigure}
    \hfill
    \begin{subfigure}[t]{0.47\linewidth}
        \centering
        \includegraphics[width=\linewidth]{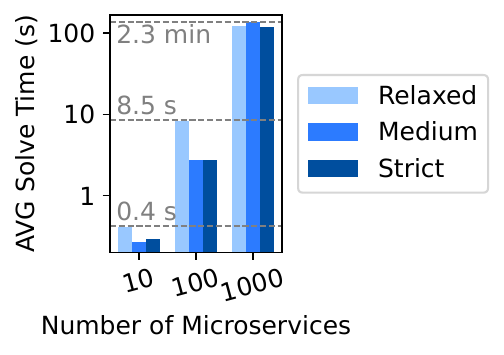}
        \vspace{-17pt}
        \caption{Solve time vs. microservice count and SLO strictness.}
        \label{fig:time-scale-slo}
    \end{subfigure}
    \caption{Scalability of \sys. 
    }
    \label{fig:scalability}
    \vspace{-0.1in}
\end{figure}

\subsection{Scalability}
\label{sec:eval_scalability}

Next, we evaluate \sys's scalability as application size and candidate region count 
increase. 
Figure~\ref{fig:time-scale} shows solve time vs the number of microservices and groups of regions (EU, EU+US, All).
Even with 1,000 microservices and all global regions, \sys converges in under 2.2 minutes, and takes seconds for small deployments. 

Next, we evaluate \sys's ability to scale as the optimization problem becomes harder to solve, which occurs under stricter SLO latencies. Figure~\ref{fig:time-scale-slo} reports solve time under three SLO scales: \emph{relaxed} (no effective SLO), \emph{medium} and \emph{strict} (latency reaches 80\% and 90\% of the SLO, following~\cite{parties}). With 100 microservices, solve time decreases under stricter SLOs because \sys prunes more infeasible placements. With 1,000 microservices, solve time is largely unaffected, as complexity is dominated by the number of candidate region-microservice mappings even after pruning.

These results show that \sys handles 
typical deployment scales (Section~\ref{sec:back_ms}) within seconds. 
Given that workload bursts last 15–45 minutes~\cite{AzureTrace, ms-oscillation} and carbon shifts occur at coarse diurnal intervals~\cite{electricity-map}, optimization is triggered only every few hours. 
\sys consistently solves placements well within these intervals, maintaining responsiveness and adaptability.

\subsection{Ablation Study}
\label{sec:eval_ablation}

\subsubsection{Optimizer}

 \begin{figure}[t] 
    \centering
    \includegraphics[width=0.90\linewidth]{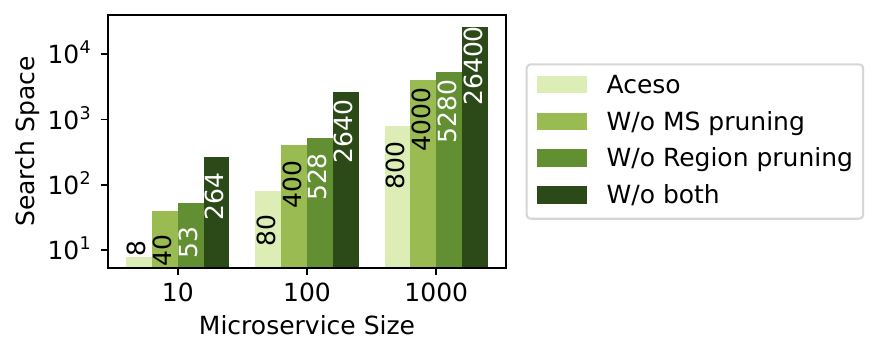}
    \vspace{-0.1in}
    \caption{Ablation study of the Placement Optimizer.}
    \label{fig:ablation-optimizer}
    \vspace{-0.1in}
\end{figure}

To assess the impact of search space pruning, we perform an ablation study of the Placement Optimizer, comparing \sys against three variants: without region filtering, without microservice pinning, and without both. Figure~\ref{fig:ablation-optimizer} shows the search space size (in log scale). 
\sys achieves the smallest space, while removing either technique increases it significantly, and disabling both leads to {\it almost two orders of magnitude larger search space}. This explains why insight-based pruning of both regions and microservices is key to scalability, enabling fast convergence even for larger deployments (Section~\ref{sec:eval_scalability}).

\subsubsection{Forecaster}
To evaluate the impact of the Traffic Forecaster, we repeat our baseline experiment (Section~\ref{sec:eval_baselines}) while disabling the predictor. In this \textit{reactive mode}, 
workload changes directly trigger placement updates, yielding negligible differences in carbon, cost, and latency (within $\pm$1\%). However, the absence of forecasting substantially decreases system stability.
Optimization triggers increase from 27 to 113, and only 71.6\% lead to adaptations, 
resulting in a $4\times$ increase in placement changes (28--29 vs.\ 6--7 per day).

 \begin{figure}[t] 
    \centering
    \includegraphics[width=0.99\linewidth]{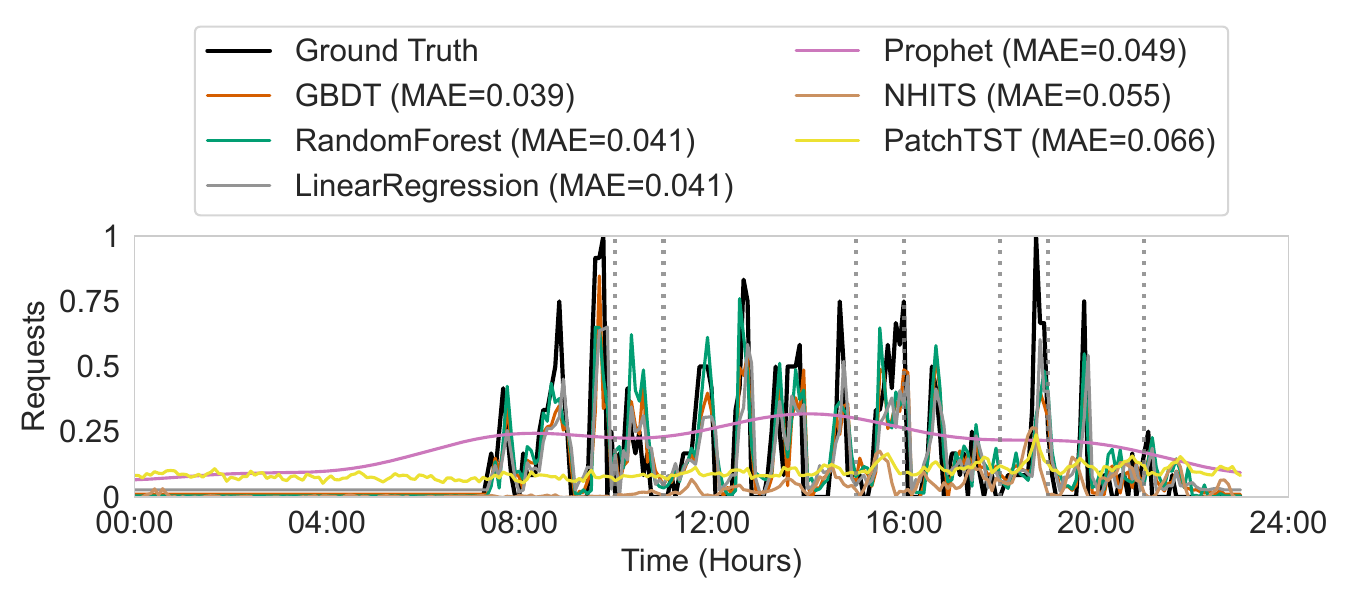}
    \vspace{-0.1in}
    \caption{Traffic prediction over 24 hours using different models; optimization triggers are marked with vertical lines.
    }
    \label{fig:eval_forecaster}
    \vspace{-0.2in}
\end{figure}

\vspace{0.05in}
\noindent{\bf Evaluation of the Traffic Forecaster.}
\label{sec:eval_ml}
We evaluate traffic prediction using the Microsoft Azure Functions Invocation Trace~\cite{AzureTrace}, widely used in prior work~\cite{ecolife,caribou,scaling-batch,microservices-azure,usher-azure,m-serve-azure}. The dataset records function invocations at five-minute intervals and is split into 50\% training, 21\% validation, and 29\% testing (7, 3, and 4 days, respectively).
We consider a representative set of forecasting models: linear (Linear Regression), tree-based (Random Forest~\cite{random-forests}, GBDT~\cite{gradient-boosted}), statistical (Prophet~\cite{prophet, forecasting-scale-fb}), and neural (N-HiTS~\cite{nhits}, PatchTST~\cite{patchtst}).
These models are widely used to predict resource usage~\cite{kerveros, barista, astraea, random-forest-gpu}, guide scheduling decisions~\cite{lava-gbdt, statuscale}, and support performance management~\cite{how-does-it-function, ml-performance-prediction-nsdi, atc-primo-models}.

GBDT achieves the lowest MAE with sub-millisecond inference, outperforming all baselines, as shown in Figure~\ref{fig:eval_forecaster}. Random Forest and Linear Regression follow closely but underpredict traffic spikes. Prophet captures trends but misses short-term fluctuations, leading to errors during peak times. Neural models (N-HiTS, PatchTST) underperform as they target long-horizon forecasting and require large histories, which are unavailable and costly to maintain. 
These results align with prior work showing that lightweight models are sufficient for short-horizon traffic forecasting~\cite{imdea-is-ml-necessary, seagull, casper, take-it-to-the-limit, deep-or-stat, amazon-intelligent-scaling, europar-christofidi, overcommitment-imdea, imdea-lstms}. Overall, GBDT offers the best balance of accuracy and efficiency for real-time forecasting in \sys.

\subsection{Sensitivity Analysis}
\label{sec:eval_sensitivity}

\begin{figure}[t]
    \centering
    \begin{subfigure}[t]{0.4\linewidth}
        \centering
        \includegraphics[width=\linewidth]{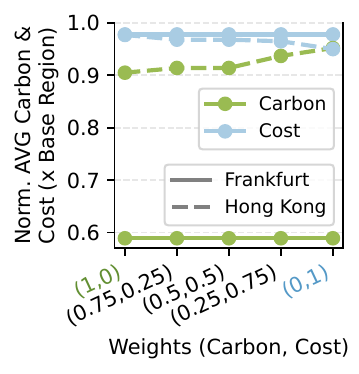}
        \vspace{-14pt}
        \caption{Weight sensitivity.
        }
        \label{fig:sensitiv}
    \end{subfigure}
    \hfill
    \begin{subfigure}[t]{0.59\linewidth}
        \centering
        \includegraphics[width=0.8\linewidth]{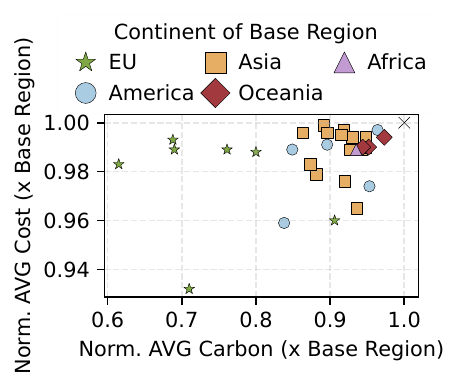}
        \vspace{-5pt}
        \caption{Generalizability. 
        }
        \label{fig:Generalizability}
    \end{subfigure}
    \caption{Sensitivity to the weight selection and geographic generalizability of \sys. }
    \label{fig:sensitivity-generalizability}
    \vspace{-0.2in}
\end{figure}

Application owners can configure \sys with different weights to prioritize carbon or cost. Figure~\ref{fig:sensitiv} illustrates the impact of weight selection on both metrics using Frankfurt (EU) and Hong Kong (Asia) as base regions for an application with 100 microservices. 
In Europe, both metrics remain largely unchanged across all weight combinations (41.1\% carbon savings and 2.2\% cost savings), 
indicating that the optimization is largely insensitive to weight selection. This happens because in Europe regions like Sweden provide substantial carbon reductions while also being among the lowest-cost options. As a result, the optimizer converges to the same geo-distribution regardless of the chosen weights.
In Asia, the differences in carbon and cost savings are more pronounced (4.8\%-9.5\% carbon savings and 2.3\%-5\% cost savings), with savings increasing as the corresponding metric is prioritized. This happens because in Asia, the carbon and cost profiles of the regions are more diverse (Figures~\ref{fig:CI_combined} and~\ref{fig:region_cost}). For example, Kuala Lumpur exhibits significantly higher carbon intensity but substantially lower cost compared to Auckland. Therefore, changing weights leads to a different selection of regions and microservice geo-distribution, impacting carbon and cost reduction. 
Overall, while application owners can tune weights to explore different cost–carbon trade-offs, by default \sys uses equal weights (0.5, 0.5), providing a balanced joint optimization of both objectives.

\subsection{Geographic Generalizability}
\label{sec:eval_generalizability}

This section analyzes how the carbon and cost reductions of \sys vary across different base locations worldwide for an application of 100 microservices. Figure~\ref{fig:Generalizability} shows carbon  
and cost,  
normalized to the baseline where all microservices are placed in the base region, across AWS regions. 
We observe that base locations in the EU achieve the largest gains, reaching up to 38.5\% carbon reduction (Frankfurt) and 6.8\% cost reduction (Zurich). This is due to the presence of multiple nearby low-carbon regions (e.g., Stockholm, Paris), which allow a large fraction of microservices to be offloaded while respecting latency constraints. 
In contrast, regions in the rest of the world exhibit more moderate {\it relative} carbon and cost reductions. However, in absolute terms, an 8.5\% reduction in Thailand corresponds to a decrease of 40.8\,gCO$_2$e/kWh, while a 31.2\% reduction in London to 49.3\,gCO$_2$e/kWh~\cite{electricity-map}. This shows that regions with higher baseline carbon intensity can still achieve substantial absolute gains, even when latency-aware geo-distribution is more limited due to larger inter-region distances. Cost savings are more modest, reflecting the narrow price differences across AWS regions (Figure 2), but they can accumulate at scale and over time.
Notably, for Sweden, N.\ Virginia, Ohio, and Oregon (marked with $\times$), no geo-distribution happens when optimizing for both carbon and cost, due to them already having the lowest carbon intensity (Sweden) or lowest cost (the latter three). 
Overall, \sys achieves up to 38.5\% (12.1\% on average) carbon reduction and up to 6.8\% (1.5\% on average) cost reduction across regions, demonstrating its ability to generalize and provide substantial gains across diverse geographic settings.

\section{Related Work}
\label{sec:related}

Table~\ref{tab:related_works} summarizes the most relevant studies and highlights how our approach differs from existing work.
Several works address carbon-aware autoscaling. CarbonScaler~\cite{carbon-scaler} adjusts server capacity for batch workloads based on carbon forecasts, while CarbonEdge~\cite{carbonedge} relocates applications to greener regions. GAIA~\cite{green-for-less-green} and Caribou~\cite{caribou} target batch and serverless workloads, respectively, incorporating cost-awareness to encourage adoption. Our work follows a similar rationale, aiming to reduce both carbon emissions and operational costs.
In the serverless domain, Ecolife~\cite{ecolife} aligns execution with low-carbon periods, while Nautilus~\cite{nautilus} and Oakestra~\cite{oakestra} enable distributed microservice orchestration without considering carbon or cost. \camready{Load-balancing systems such as Nautilus~\cite{nautilus}, FIRM~\cite{firm}, and Imbres~\cite{imbres} are also not carbon-aware, they optimize performance within a region,} and are complementary to our work. \camready{Temporal shifting approaches (e.g., Let’s Wait a While~\cite{lets-wait-awhile}, S.C.A.L.E.~\cite{Toonder2024}, PCAPS~\cite{Lechowicz2025} for delay-tolerant, sequential tasks) 
schedule workloads during cleaner periods, but are less effective for latency-sensitive applications, since they can introduce hour-scale delays until greener energy becomes available}, and as grids decarbonize~\cite{on-the-limitations}.  
Complementary work includes frameworks such as CarbonExplorer~\cite{carbonexplorer} for datacenter design and GreenSKUs~\cite{greensku} for low-carbon server configurations. In contrast, \sys targets SMEs on commodity cloud infrastructures. The Sunk Carbon Fallacy~\cite{sunk-carbon} further highlights the importance of accounting for embodied carbon in system design.

\begin{table}[t]
    \small
    \centering
    \renewcommand\cellalign{lc}
    \resizebox{\columnwidth}{!}{%
    \begin{tabular}{p{4cm}p{0.4cm}p{0.9cm}p{0.8cm}p{0.8cm}} \hline
         \makecell{\bf Solutions} &
         \makecell{\bf MS \\ \bf App} &
         \makecell{\bf Latency \\ \bf Compl.} &
         \makecell{\bf Carbon \\ \bf Aware} &
         \makecell{\bf Cost \\ \bf Efficient} \\\hline
         \makecell{CarbonScaler~\cite{carbon-scaler}, Let's Wait \\ a While~\cite{lets-wait-awhile}, S.C.A.L.E\cite{Toonder2024}} & \xmark & \xmark & \cmark~~ & \xmark  \\\hline
         \makecell{CarbonEdge~\cite{carbonedge}, Ecolife~\cite{ecolife}, \\ KEIDS~\cite{hadidi2023keids}}& \xmark & \cmark & \cmark~~ & \xmark \\\hline
         GAIA~\cite{green-for-less-green} & \xmark & \xmark & \cmark~~ & \cmark \\\hline
         \makecell{Oakestra~\cite{oakestra}, Astraea~\cite{astraea}, \\ Nautilus~\cite{nautilus}, FIRM~\cite{firm},\\ Imbres~\cite{imbres}} & \cmark & \cmark & \xmark & \xmark \\\hline
         Caribou~\cite{caribou} & \xmark & \cmark & \cmark~~ & \cmark \\\hline
         NSGA-II Opt~\cite{cortellessa-ga} & \cmark & \cmark & \xmark & \cmark \\
         \hline
         PCAPS\cite{Lechowicz2025}& \cmark & \xmark & \cmark & \xmark \\\hline
         CASPER~\cite{wolski2023casper}&\cmark & \cmark & \cmark & \xmark \\\hline\hline
         {\bf \sys} & \cmark & \cmark & \cmark~~ & \cmark \\\hline
    \end{tabular}
    }
        \caption{Comparison of frameworks related to \sys. 
        }
    \label{tab:related_works}
    \vspace{-0.3in}
\end{table}

\section{Discussion \& Future Work}
\label{sec:discussion}

\noindent{\bf Current limitations of \sys.} 
First, the optimizer depends on real-time carbon-intensity and pricing signals, which may occasionally be stale or delayed, affecting decision accuracy despite our local caching mechanisms.
Second, our evaluation focuses on AWS-based deployments; generalizing to other cloud providers, or hybrid infrastructures may introduce new challenges and opportunities, such as region availability, inter-region latencies, data transfer policies, or region-specific regulatory constraints. 
For example, AWS has limited regional coverage in some areas (e.g., only São Paulo in South America), whereas Google Cloud offers additional nearby regions (e.g., São Paulo and Santiago~\cite{google-regions}), enabling more flexible deployments and potentially higher improvements. 

\noindent{\bf Future Work.}
Future research will extend \sys along multiple directions.
We plan to:
1)~explore integration with 
spot pricing to further reduce costs;
2)~enhance the system with tiered SLOs, to enable differentiated service levels for premium and best-effort users and tasks;
3)~develop a dashboard for real-time monitoring and interactive configuration, to improve usability for SMEs.

\section{Summary}
This paper presents \sys, a system that dynamically places microservices across geographically constrained settings to reduce carbon emissions in a cost-efficient way while meeting latency constraints. At its core \sys uses a scalable optimization approach that leverages an insight-based pruned search space. 
Evaluation across diverse regions and application types shows that \sys achieves significant carbon and cost savings under the latency targets of microservice applications.

\begin{acks}
        We thank the anonymous reviewers for their constructive feedback and our shepherd Li (Lilly) Wu for her guidance.
        The work by the authors at the IMDEA Software Institute was partially funded by the Madrid Regional Government through the César Nombela grant (2024-T1/COM-31302) and by the Comunidad de Madrid through the DATIA project, co-funded by the European Union's FEDER funds. Their work was also supported by grant PID2022-142290OB-I00, funded by the Spanish Ministry of Science, Innovation and Universities (MCIN/AEI/10.13039/501100011033) and FEDER, UE, and by grant CEX2024-001471-M, also funded by the Spanish Ministry of Science, Innovation and Universities (MCIN/AEI/10.13039/ 501100011033).
        The work by J.O.I is partially supported by the EU EDGELESS project funded by the EU programme (agreement No. 101092950) and by the Smart Networks and Services Joint Undertaking (SNS JU) under the EU HE programme (agreement No. 101293102).
\end{acks}

\balance

\bibliographystyle{ACM-Reference-Format}
\bibliography{paper}

@article{pedregosa2011scikit,
  title={Scikit-learn: Machine Learning in Python},
  author={Pedregosa, Fabian and Varoquaux, Ga{\"e}l and Gramfort, Alexandre and Michel, Vincent and Thirion, Bertrand and Grisel, Olivier and Blondel, Mathieu and Prettenhofer, Peter and Weiss, Ron and Dubourg, Vincent and others},
  journal={Journal of Machine Learning Research},
  volume={12},
  pages={2825--2830},
  year={2011}
}

@misc{garza2022neuralforecast,
  title={NeuralForecast: Deep Learning for Time Series Forecasting},
  author={Garza, Alejandro and others},
  year={2022},
  howpublished={\url{https://github.com/Nixtla/neuralforecast}}
}

@inproceedings{hadidi2023keids,
  author    = {Kaur, Kuljeet and Garg, Sahil and Kaddoum, Georges and Ahmed, Syed Hassan and Atiquzzaman, Mohammed},
  title     = {KEIDS: Kubernetes-Based Energy and Interference Driven Scheduler for Industrial IoT in Edge-Cloud Ecosystem},
  booktitle = {IEEE Internet of Things Journal},
  year      = {2020},
  pages     = {4228-4237},
  doi       = {10.1109/JIOT.2019.2939534}
}

@inproceedings{AzureTrace,
  author    = {Yanqi Zhang and Íñigo Goiri and Rodrigo Fonseca and Sameh Elnikety and Christina Delimitrou and Ricardo Bianchini},
  title     = {Faster and Cheaper Serverless Computing on Harvested Resources},
  booktitle = {Proceedings of the ACM SIGOPS 28th Symposium on Operating Systems Principles (SOSP)},
  year      = {2021},
  pages     = {724--739}
}

@inproceedings{Toonder2024,
author = {Den Toonder, Jurriaan and Braakman, Paul and Durieux, Thomas},
title = {S.C.A.L.E: A CO2-Aware Scheduler for OpenShift at ING},
year = {2024},
isbn = {9798400706585},
publisher = {Association for Computing Machinery},
address = {New York, NY, USA},
url = {https://doi.org/10.1145/3663529.3663862},
doi = {10.1145/3663529.3663862},
booktitle = {Companion Proceedings of the 32nd ACM International Conference on the Foundations of Software Engineering},
pages = {429–439},
numpages = {11},
location = {Porto de Galinhas, Brazil},
series = {FSE 2024}
}

@article{bostandoost2024data,
  title={Data-driven Algorithm Selection for Carbon-Aware Scheduling},
  author={Bostandoost, Roozbeh and Hanafy, Walid A and Lechowicz, Adam and Bashir, Noman and Shenoy, Prashant and Hajiesmaili, Mohammad},
  journal={ACM SIGENERGY Energy Informatics Review},
  volume={4},
  number={5},
  pages={148--153},
  year={2024},
  publisher={ACM New York, NY, USA}
}

@inproceedings{maji2023bringing,
  title={Bringing carbon awareness to multi-cloud application delivery},
  author={Maji, Diptyaroop and Pfaff, Ben and PR, Vipin and Sreenivasan, Rajagopal and Firoiu, Victor and Iyer, Sreeram and Josephson, Colleen and Pan, Zhelong and Sitaraman, Ramesh K},
  booktitle={Proceedings of the 2nd Workshop on Sustainable Computer Systems},
  pages={1--6},
  year={2023}
}

@inproceedings{xu2025green,
  title={$\{$GREEN$\}$: Carbon-efficient Resource Scheduling for Machine Learning Clusters},
  author={Xu, Kaiqiang and Sun, Decang and Tian, Han and Zhang, Junxue and Chen, Kai},
  booktitle={22nd USENIX Symposium on Networked Systems Design and Implementation (NSDI 25)},
  pages={999--1014},
  year={2025}
}

@inproceedings{wolski2023casper,
  title={Casper: Carbon-aware scheduling and provisioning for distributed web services},
  author={Souza, Abel and Jasoria, Shruti and Chakrabarty, Basundhara and Bridgwater, Alexander and Lundberg, Axel and Skogh, Filip and Ali-Eldin, Ahmed and Irwin, David and Shenoy, Prashant},
  booktitle={Proceedings of the 14th International Green and Sustainable Computing Conference},
  pages={67--73},
  year={2023}
}

@article{ice-buckets,
  title={Ice buckets: Improved counter estimation for network measurement},
  author={Einziger, Gil and Fellman, Benny and Friedman, Roy and Kassner, Yaron},
  journal={IEEE/ACM Transactions on Networking},
  volume={26},
  number={3},
  pages={1165--1178},
  year={2018},
  publisher={IEEE}
}

@inproceedings{Lechowicz2025,
author = {Lechowicz, Adam and Shenoy, Rohan and Bashir, Noman and Hajiesmaili, Mohammad and Wierman, Adam and Delimitrou, Christina},
title = {Carbon- and Precedence-Aware Scheduling for Data Processing Clusters},
year = {2025},
isbn = {9798400715242},
publisher = {Association for Computing Machinery},
address = {New York, NY, USA},
url = {https://doi.org/10.1145/3718958.3750478},
doi = {10.1145/3718958.3750478},
booktitle = {Proceedings of the ACM SIGCOMM 2025 Conference},
pages = {1241–1244},
numpages = {4},
location = {S\~{a}o Francisco Convent, Coimbra, Portugal},
series = {SIGCOMM '25}
}

@misc{un_sdg_2015,
  title        = {Transforming our world: the 2030 Agenda for Sustainable Development},
  author       = {{United Nations}},
  year         = {2015},
  howpublished = {\url{https://sdgs.un.org/goals}},
  note         = {Adopted by all United Nations Member States in 2015},
}

@misc{green_procurement_ec2025,
  title        = {Green Public Procurement},
  author       = {{European Commission}},
  howpublished = {\url{https://green-forum.ec.europa.eu/green-business/green-public-procurement_en}},
}

@misc{eu_ets_2003,
  title        = {Directive 2003/87/EC establishing a scheme for greenhouse gas emission allowance trading within the Community},
  author       = {{European Parliament and Council}},
  year         = {2003},
  howpublished = {\url{https://eur-lex.europa.eu/eli/dir/2003/87/oj/eng}},
  note         = {OJ L 275, 25.10.2003, pp. 32–46}
}

@techreport{carbon_tax_jrc2024,
  author       = {Maier, Sofia and De Poli, Silvia and Amores, Antonio F.},
  title        = {Carbon taxes on consumption: distributional implications for a just transition in the EU},
  institution  = {European Commission, Joint Research Centre},
  year         = {2024},
  number       = {JRC138420},
  url          = {https://publications.jrc.ec.europa.eu/repository/handle/JRC138420},
  note         = {JRC Working Papers on Taxation and Structural Reforms No 9/2024}
}

@ARTICLE{ms-placement-joined-opt,
  author={Peng, Kai and Wang, Liangyuan and He, Jintao and Cai, Chao and Hu, Menglan},
  journal={IEEE Transactions on Services Computing}, 
  title={Joint Optimization of Service Deployment and Request Routing for Microservices in Mobile Edge Computing}, 
  year={2024},
  volume={17},
  number={3},
  pages={1016-1028},
  doi={10.1109/TSC.2024.3349408}}

@misc{europeanclimatelaw2021,
  title        = {Regulation (EU) 2021/1119 of the European Parliament and of the Council of 30 June 2021 establishing the framework for achieving climate neutrality and amending Regulations (EC) No 401/2009 and (EU) 2018/1999 (‘European Climate Law’)},
  author       = {{European Parliament and Council}},
  year         = {2021},
  month        = {July},
  howpublished = {\url{https://eur-lex.europa.eu/legal-content/EN/TXT/?uri=CELEX:32021R1119}},
  note         = {OJ L 243, 9.7.2021, pp. 1–17},
}

@misc{eugreendeal2019,
  title        = {The European Green Deal},
  author       = {{European Commission}},
  year         = {2019},
  month        = {December},
  howpublished = {\url{https://eur-lex.europa.eu/legal-content/EN/TXT/?uri=CELEX:52019DC0640}},
  note         = {COM(2019) 640 final},
}

@article{fastpert, title={FastPERT: Towards Fast Microservice Application Latency Prediction via Structural Inductive Bias over PERT Networks}, volume={39}, url={https://ojs.aaai.org/index.php/AAAI/article/view/34291}, DOI={10.1609/aaai.v39i19.34291}, abstractNote={The recent surge in popularity of cloud-native applications using microservice architectures has led to a focus on accurate end-to-end latency prediction for proactive resource allocation. Existing models leverage Graph Transformers to Microservice Call Graphs or the Program Evaluation and Review Technique (PERT) graphs to capture complex temporal dependencies between microservices. However, these models incur a high computational cost during both training and inference phases. This paper introduces FastPERT, an efficient model for predicting end-to-end latency in microservice applications. FastPERT dissects an execution trace into several microservices tasks, using observations from prior execution traces of the application, akin to the PERT approach. Subsequently, a prediction model is constructed to estimate the completion time for each individual task. This information, coupled with the computational and structural inductive bias of the PERT graph, facilitates the efficient computation of the end-to-end latency of an execution trace. As a result, FastPERT can efficiently capture the complex temporal causality of different microservice tasks without relying on Graph Neural Networks, leading to more accurate and robust latency predictions across a variety of applications. An evaluation based on datasets generated from large-scale Alibaba microservice traces reveals that FastPERT significantly improves training and inference efficiency without compromising performance, demonstrating its potential as a superior solution for real-time end-to-end latency prediction in cloud-native microservice applications.}, number={19}, journal={Proceedings of the AAAI Conference on Artificial Intelligence}, author={Tam, Da Sun Handason and Xu, Huanle and Liu, Yang and Xie, Siyue and Lau, Wing Cheong}, year={2025}, month={Apr.}, pages={20787–20795} }

@inproceedings{graf,
author = {Park, Jinwoo and Choi, Byungkwon and Lee, Chunghan and Han, Dongsu},
title = {GRAF: a graph neural network based proactive resource allocation framework for SLO-oriented microservices},
year = {2021},
isbn = {9781450390989},
publisher = {Association for Computing Machinery},
address = {New York, NY, USA},
url = {https://doi.org/10.1145/3485983.3494866},
doi = {10.1145/3485983.3494866},
booktitle = {Proceedings of the 17th International Conference on Emerging Networking EXperiments and Technologies},
pages = {154–167},
numpages = {14},
location = {Virtual Event, Germany},
series = {CoNEXT '21}
}

@ARTICLE{call-graphs-ieee,
  author={Hu, Yi and Wang, Hao and Wang, Liangyuan and Hu, Menglan and Peng, Kai and Veeravalli, Bharadwaj},
  journal={IEEE Transactions on Parallel and Distributed Systems}, 
  title={Joint Deployment and Request Routing for Microservice Call Graphs in Data Centers}, 
  year={2023},
  volume={34},
  number={11},
  pages={2994-3011},
  doi={10.1109/TPDS.2023.3311767}}

@article{Spatiotemporal-online,
author = {Lechowicz, Adam and Christianson, Nicolas and Sun, Bo and Bashir, Noman and Hajiesmaili, Mohammad and Wierman, Adam and Shenoy, Prashant},
title = {Learning-Augmented Competitive Algorithms for Spatiotemporal Online Allocation with Deadline Constraints},
year = {2025},
issue_date = {March 2025},
publisher = {Association for Computing Machinery},
address = {New York, NY, USA},
volume = {9},
number = {1},
url = {https://doi.org/10.1145/3711701},
doi = {10.1145/3711701},
journal = {Proc. ACM Meas. Anal. Comput. Syst.},
month = mar,
articleno = {8},
numpages = {49}
}

@Article{iot-ms-placement,
AUTHOR = {Canali, Claudia and Gazzotti, Caterina and Lancellotti, Riccardo and Schena, Felice},
TITLE = {Placement of IoT Microservices in Fog Computing Systems: A Comparison of Heuristics},
JOURNAL = {Algorithms},
VOLUME = {16},
YEAR = {2023},
NUMBER = {9},
ARTICLE-NUMBER = {441},
URL = {https://www.mdpi.com/1999-4893/16/9/441},
ISSN = {1999-4893},
DOI = {10.3390/a16090441}
}

@inproceedings{sunk-carbon,
author = {Bashir, Noman and Gohil, Varun and Subramanya, Anagha Belavadi and Shahrad, Mohammad and Irwin, David and Olivetti, Elsa and Delimitrou, Christina},
title = {The Sunk Carbon Fallacy: Rethinking Carbon Footprint Metrics for Effective Carbon-Aware Scheduling},
year = {2024},
isbn = {9798400712869},
publisher = {Association for Computing Machinery},
address = {New York, NY, USA},
url = {https://doi.org/10.1145/3698038.3698542},
doi = {10.1145/3698038.3698542},
booktitle = {Proceedings of the 2024 ACM Symposium on Cloud Computing},
pages = {542–551},
numpages = {10},
location = {Redmond, WA, USA},
series = {SoCC '24}
}

@inproceedings{carboncast,
author = {Maji, Diptyaroop and Shenoy, Prashant and Sitaraman, Ramesh K.},
title = {CarbonCast: multi-day forecasting of grid carbon intensity},
year = {2022},
isbn = {9781450398909},
publisher = {Association for Computing Machinery},
address = {New York, NY, USA},
url = {https://doi.org/10.1145/3563357.3564079},
doi = {10.1145/3563357.3564079},
booktitle = {Proceedings of the 9th ACM International Conference on Systems for Energy-Efficient Buildings, Cities, and Transportation},
pages = {198–207},
numpages = {10},
location = {Boston, Massachusetts},
series = {BuildSys '22}
}

@inproceedings{on-the-limitations,
author = {Sukprasert, Thanathorn and Souza, Abel and Bashir, Noman and Irwin, David and Shenoy, Prashant},
title = {On the Limitations of Carbon-Aware Temporal and Spatial Workload Shifting in the Cloud},
year = {2024},
isbn = {9798400704376},
publisher = {Association for Computing Machinery},
address = {New York, NY, USA},
url = {https://doi.org/10.1145/3627703.3650079},
doi = {10.1145/3627703.3650079},
booktitle = {Proceedings of the Nineteenth European Conference on Computer Systems},
pages = {924–941},
numpages = {18},
location = {Athens, Greece},
series = {EuroSys '24}
}

@INPROCEEDINGS{greensku,
  author={Wang, Jaylen and Berger, Daniel S. and Kazhamiaka, Fiodar and Irvene, Celine and Zhang, Chaojie and Choukse, Esha and Frost, Kali and Fonseca, Rodrigo and Warrier, Brijesh and Bansal, Chetan and Stern, Jonathan and Bianchini, Ricardo and Sriraman, Akshitha},
  booktitle={2024 ACM/IEEE 51st Annual International Symposium on Computer Architecture (ISCA)}, 
  title={Designing Cloud Servers for Lower Carbon}, 
  year={2024},
  volume={},
  number={},
  pages={452-470},
  doi={10.1109/ISCA59077.2024.00041}}

@misc{carbonedge,
      title={CarbonEdge: Leveraging Mesoscale Spatial Carbon-Intensity Variations for Low Carbon Edge Computing}, 
      author={Li Wu and Walid A. Hanafy and Abel Souza and Khai Nguyen and Jan Harkes and David Irwin and Mahadev Satyanarayanan and Prashant Shenoy},
      year={2025},
      eprint={2502.14076},
      archivePrefix={arXiv},
      primaryClass={cs.DC},
      url={https://arxiv.org/abs/2502.14076}, 
}

@inproceedings{carbonexplorer,
author = {Acun, Bilge and Lee, Benjamin and Kazhamiaka, Fiodar and Maeng, Kiwan and Gupta, Udit and Chakkaravarthy, Manoj and Brooks, David and Wu, Carole-Jean},
title = {Carbon Explorer: A Holistic Framework for Designing Carbon Aware Datacenters},
year = {2023},
isbn = {9781450399166},
publisher = {Association for Computing Machinery},
address = {New York, NY, USA},
url = {https://doi.org/10.1145/3575693.3575754},
doi = {10.1145/3575693.3575754},
booktitle = {Proceedings of the 28th ACM International Conference on Architectural Support for Programming Languages and Operating Systems, Volume 2},
pages = {118–132},
numpages = {15},
location = {Vancouver, BC, Canada},
series = {ASPLOS 2023}
}

@inproceedings{ecolife,
author = {Jiang, Yankai and Roy, Rohan Basu and Li, Baolin and Tiwari, Devesh},
title = {EcoLife: Carbon-Aware Serverless Function Scheduling for Sustainable Computing},
year = {2024},
isbn = {9798350352917},
publisher = {IEEE Press},
url = {https://doi.org/10.1109/SC41406.2024.00018},
doi = {10.1109/SC41406.2024.00018},
booktitle = {Proceedings of the International Conference for High Performance Computing, Networking, Storage, and Analysis},
articleno = {12},
numpages = {15},
location = {Atlanta, GA, USA},
series = {SC '24}
}

@article{ga-app-placement,
title = {Distributed genetic algorithm for application placement in the compute continuum leveraging infrastructure nodes for optimization},
journal = {Future Generation Computer Systems},
volume = {160},
pages = {154-170},
year = {2024},
issn = {0167-739X},
doi = {https://doi.org/10.1016/j.future.2024.05.044},
url = {https://www.sciencedirect.com/science/article/pii/S0167739X24002760},
author = {Carlos Guerrero and Isaac Lera and Carlos Juiz}
}

@inproceedings{feng2017approximation,
  title={Approximation algorithms for the NFV service distribution problem},
  author={Feng, Hao and Llorca, Jaime and Tulino, Antonia M and Raz, Danny and Molisch, Andreas F},
  booktitle={IEEE INFOCOM 2017-IEEE Conference on Computer Communications},
  pages={1--9},
  year={2017},
  organization={IEEE}
}

@article{baev2008approximation,
  title={Approximation algorithms for data placement problems},
  author={Baev, Ivan and Rajaraman, Rajmohan and Swamy, Chaitanya},
  journal={SIAM Journal on Computing},
  volume={38},
  number={4},
  pages={1411--1429},
  year={2008},
  publisher={SIAM}
}

@article{yang2019delay,
  title={Delay-aware virtual network function placement and routing in edge clouds},
  author={Yang, Song and Li, Fan and Trajanovski, Stojan and Chen, Xu and Wang, Yu and Fu, Xiaoming},
  journal={IEEE Transactions on Mobile Computing},
  volume={20},
  number={2},
  pages={445--459},
  year={2019},
  publisher={IEEE}
}

@inproceedings{parties,
author = {Chen, Shuang and Delimitrou, Christina and Mart\'{\i}nez, Jos\'{e} F.},
title = {PARTIES: QoS-Aware Resource Partitioning for Multiple Interactive Services},
year = {2019},
isbn = {9781450362405},
publisher = {Association for Computing Machinery},
address = {New York, NY, USA},
url = {https://doi.org/10.1145/3297858.3304005},
doi = {10.1145/3297858.3304005},
booktitle = {Proceedings of the Twenty-Fourth International Conference on Architectural Support for Programming Languages and Operating Systems},
pages = {107–120},
numpages = {14},
location = {Providence, RI, USA},
series = {ASPLOS '19}
}

@inproceedings{astraea,
author = {Zhang, Wei and Chen, Quan and Fu, Kaihua and Zheng, Ningxin and Huang, Zhiyi and Leng, Jingwen and Guo, Minyi},
title = {Astraea: towards QoS-aware and resource-efficient multi-stage GPU services},
year = {2022},
isbn = {9781450392051},
publisher = {Association for Computing Machinery},
address = {New York, NY, USA},
url = {https://doi.org/10.1145/3503222.3507721},
doi = {10.1145/3503222.3507721},
booktitle = {Proceedings of the 27th ACM International Conference on Architectural Support for Programming Languages and Operating Systems},
pages = {570–582},
numpages = {13},
location = {Lausanne, Switzerland},
series = {ASPLOS '22}
}

@ARTICLE{nautilus,
  author={Fu, Kaihua and Zhang, Wei and Chen, Quan and Zeng, Deze and Guo, Minyi},
  journal={IEEE Transactions on Parallel and Distributed Systems}, 
  title={Adaptive Resource Efficient Microservice Deployment in Cloud-Edge Continuum}, 
  year={2022},
  volume={33},
  number={8},
  pages={1825-1840},
  doi={10.1109/TPDS.2021.3128037}}

@inproceedings{lets-wait-awhile,
author = {Wiesner, Philipp and Behnke, Ilja and Scheinert, Dominik and Gontarska, Kordian and Thamsen, Lauritz},
title = {Let's wait awhile: how temporal workload shifting can reduce carbon emissions in the cloud},
year = {2021},
isbn = {9781450385343},
publisher = {Association for Computing Machinery},
address = {New York, NY, USA},
url = {https://doi.org/10.1145/3464298.3493399},
doi = {10.1145/3464298.3493399},
booktitle = {Proceedings of the 22nd International Middleware Conference},
pages = {260–272},
numpages = {13},
location = {Qu\'{e}bec city, Canada},
series = {Middleware '21}
}

@article{sustainable-ai-meta,
  title={Sustainable ai: Environmental implications, challenges and opportunities},
  author={Wu, Carole-Jean and Raghavendra, Ramya and Gupta, Udit and Acun, Bilge and Ardalani, Newsha and Maeng, Kiwan and Chang, Gloria and Aga, Fiona and Huang, Jinshi and Bai, Charles and others},
  journal={Proceedings of Machine Learning and Systems},
  volume={4},
  pages={795--813},
  year={2022}
}

@article{forecasting-scale-fb,
author = {Sean J. Taylor and Benjamin Letham},
title = {Forecasting at Scale},
journal = {The American Statistician},
volume = {72},
number = {1},
pages = {37--45},
year = {2018},
publisher = {ASA Website},
doi = {10.1080/00031305.2017.1380080},


URL = { 
    
        https://doi.org/10.1080/00031305.2017.1380080
    
    

},
eprint = { 
    
        https://doi.org/10.1080/00031305.2017.1380080
    
    

}

}

@inproceedings{lava-gbdt,
title={{LAVA}: Lifetime-Aware {VM} Allocation with Learned Distributions and Adaptation to Mispredictions},
author={Jianheng Ling and Pratik Worah and Yawen Wang and Yunchuan Kong and Chunlei Wang and Clifford Stein and Diwakar Gupta and Jason Behmer and Logan A. Bush and Prakash Ramanan and Rajesh Kumar and Thomas Chestna and Yajing Liu and Ying Liu and Ye Zhao and Kathryn S. McKinley and Meeyoung Park and Martin Maas},
booktitle={Eighth Conference on Machine Learning and Systems},
year={2025},
url={https://openreview.net/forum?id=9vyyfVNW1E}
}

@misc{micro_techgiants,
  author       = {Akhil Sundar},
  title        = {9 Tech Giants Embracing Microservices Architecture},
  howpublished = {Blog post, SayOneTech},
  year         = {2021},
  month        = {Oct},
  day          = {19},
  note         = {\url{https://www.sayonetech.com/blog/microservices-architecture-companies/}}
}

@misc{microservices_use_cases,
  author       = {Yuriy Rybkin},
  title        = {Microservices Use Cases and Real-World Examples},
  howpublished = {Blog post, CodeIT},
  year         = {2025},
  month        = {Mar},
  day          = {7},
  note         = {\url{https://codeit.us/blog/microservices-use-cases}}
}

@misc{emizentech_microservices_architecture,
  author       = {Ganesh Tharol},
  title        = {Microservices Architecture: Benefits, Use Cases \& Examples},
  howpublished = {Blog post, EmizenTech},
  year         = {2022},
  note         = {\url{https://emizentech.com/blog/microservices-architecture.html}}
}

@misc{microservices-uber2020_doma,
  author       = {{Adam Gluck}},
  title        = {Introducing Domain-Oriented Microservice Architecture},
  howpublished = {Blog post, Uber Engineering Blog},
  year         = {2020},
  month        = {Jul},
  day          = {23},
  note         = {\url{https://www.uber.com/en-ES/blog/microservice-architecture/}}
}

@misc{automq-stateless-vs-stateful,
  author       = {{AutoMQ Team}},
  title        = {Stateless vs. Stateful Architecture: A Comprehensive Comparison},
  howpublished = {Blog post},
  year         = {2025},
  month        = {May},
  day          = {15},
  note         = {\url{https://www.automq.com/blog/stateless-vs-stateful-architecture-a-comprehensive-comparison}}
}

@misc{devopedia-stateless-vs-stateful,
  author       = {{Arvind Padmanabhan, Arjun G.}},
  title        = {Microservices},
  howpublished = {Blog post},
  year         = {2021},
  month        = {March},
  day          = {28},
  note         = {\url{https://devopedia.org/microservices}}
}

@misc{medium-stateless,
  author       = {{Sanket Panhale}},
  title        = {Part 2: Microservices Architecture and Stateless Design for Scalable Systems},
  howpublished = {Blog post},
  year         = {2024},
  month        = {Oct},
  day          = {25},
  note         = {\url{https://sanket-panhale.medium.com/part-2-microservices-architecture-and-stateless-design-for-scalable-systems-3b67aa88fc78}}
}

@misc{medium-stateless-guide,
  title        = {Microservice Best Practices: A Guide to Building Scalable and Efficient Systems},
  howpublished = {Blog post},
  year         = {2024},
  month        = {Dec},
  day          = {2},
  note         = {\url{https://noncodersuccess.medium.com/microservice-best-practices-a-guide-to-building-scalable-and-efficient-systems-644ba7dbca43}}
}

@misc{medium-stateless-success,
  title        = {Scaling Microservices: Key Architectural Decisions for Success},
  howpublished = {Blog post},
  year         = {2025},
  month        = {Feb},
  day          = {23},
  note         = {\url{https://medium.com/@CodeWithTech/scaling-microservices-key-architectural-decisions-for-success-eeae1c0391c6}}
}

@misc{medium-stateless-design,
  title        = {Stateless Design: The Backbone of Scalable Systems},
  howpublished = {Blog post},
  year         = {2025},
  month        = {Jul},
  day          = {29},
  note         ={\url{https://medium.com/@sonal.sadafal/stateless-design-the-backbone-of-scalable-systems-457be01f642a}}
}

@misc{dremio-stateless-vs-stateful,
  title        = {Microservices Architecture},
  howpublished = {Blog post},
  note         = {\url{https://www.dremio.com/wiki/microservices-architecture/}},
  year = {Accessed: 2026},
}

@misc{callista,
  title        = {Callista},
  howpublished = {\url{https://callistaenterprise.se/english/}},
  year         = {Accessed: 2026}
}

@misc{callista-git,
  title        = {Callista Enterprise - Github},
  howpublished = {\url{https://github.com/callistaenterprise/blog-microservices/tree/master/microservices}},
  year         = {Accessed: 2026}
}

@misc{bokbasen,
  title        = {Bokbasen},
  howpublished = {\url{https://www.bokbasen.no/en}},
  year         = {Accessed: 2026}
}

@misc{casavo,
  title        = {Casavo},
  howpublished = {\url{https://casavo.com/en/}},
  year         = {Accessed: 2026}
}

@misc{gdpr,
  title        = {General Data Protection Regulation - GDPR},
  howpublished = {\url{https://gdpr-info.eu}},
  year         = {Accessed: 2026}
}

@misc{bokbasen-ms,
  title        = {Bokbasen CI/CD Pipeline – Case Study},
  howpublished = {\url{https://www.orange-business.com/be-en/about-us/customer-stories/bokbasen-cicd-pipeline-case-study}},
  year         = {Accessed: 2026}
}

@misc{casavo-ms,
author = {Leo Sorge},
  title        = {Build Incremental Architecture Through Microservices with Modern Languages – Casavo’s Case Study},
  howpublished = {\url{https://www.codemotion.com/magazine/microservices/casavo-incremental-microservices/}},
  year         = {2020},
  month        = {Nov},
  day          = {12},
}

@misc{callista-ms,
  title        = {An operations model for Microservices},
  author       = {{Magnus Larsson}},
  howpublished = {\url{https://callistaenterprise.se/blogg/teknik/2015/03/25/an-operations-model-for-microservices/}},
  year         = {2015},
  month        = {March},
  day          = {25},
}

@misc{bytebytego-stateless,
  title        = {Is Microservice Architecture the Silver Bullet?},
  howpublished = {Blog post},
  note         = {\url{https://bytebytego.com/guides/is-microservice-architecture-the-silver-bullet/}},
  year = {Accessed: 2026},
}

@misc{digital-alpha-stateless,
  title        = {Stateless Microservices Accelerate Change},
  howpublished = {Blog post},
  note         = {\url{https://www.digital-alpha.com/stateless-microservices-accelerate-change/}},
  year = {Accessed: 2026},
}

@inproceedings{deepscaling,
author = {Wang, Ziliang and Zhu, Shiyi and Li, Jianguo and Jiang, Wei and Ramakrishnan, K. K. and Zheng, Yangfei and Yan, Meng and Zhang, Xiaohong and Liu, Alex X.},
title = {DeepScaling: microservices autoscaling for stable CPU utilization in large scale cloud systems},
year = {2022},
isbn = {9781450394147},
publisher = {Association for Computing Machinery},
address = {New York, NY, USA},
url = {https://doi.org/10.1145/3542929.3563469},
doi = {10.1145/3542929.3563469},
booktitle = {Proceedings of the 13th Symposium on Cloud Computing},
pages = {16–30},
numpages = {15},
location = {San Francisco, California},
series = {SoCC '22}
}

@inproceedings{deeprest,
author = {Chow, Ka-Ho and Deshpande, Umesh and Seshadri, Sangeetha and Liu, Ling},
title = {DeepRest: deep resource estimation for interactive microservices},
year = {2022},
isbn = {9781450391627},
publisher = {Association for Computing Machinery},
address = {New York, NY, USA},
url = {https://doi.org/10.1145/3492321.3519564},
doi = {10.1145/3492321.3519564},
booktitle = {Proceedings of the Seventeenth European Conference on Computer Systems},
pages = {181–198},
numpages = {18},
location = {Rennes, France},
series = {EuroSys '22}
}

@misc{ms-arch-ai,
  author       = {Meeran Malik},
  title        = {Microservices Architecture for AI Applications: Scalable Patterns and 2025 Trends},
  howpublished = {Medium blog post},
  year         = {2025},
  month        = {May},
  url          = {https://medium.com/@meeran03/microservices-architecture-for-ai-applications-scalable-patterns-and-2025-trends-5ac273eac232},
  note         = {Accessed: 2025-11-14}
}

@misc{ms-modular-ai,
  author       = {Tullie Murrell},
  title        = {Monolithic vs Modular AI Architecture: Key Trade‑Offs},
  howpublished = {Blog post on shaped.ai},
  year         = {2025},
  month        = {June},
  url          = {https://www.shaped.ai/blog/monolithic-vs-modular-ai-architecture},
  note         = {Accessed: 2025-11-14}
}

@inproceedings{tao-fb,
  title={$\{$TAO$\}$:$\{$Facebook’s$\}$ distributed data store for the social graph},
  author={Bronson, Nathan and Amsden, Zach and Cabrera, George and Chakka, Prasad and Dimov, Peter and Ding, Hui and Ferris, Jack and Giardullo, Anthony and Kulkarni, Sachin and Li, Harry and others},
  booktitle={2013 USENIX Annual Technical Conference (USENIX ATC 13)},
  pages={49--60},
  year={2013}
}

@inproceedings{uber-crisp,
  title={$\{$CRISP$\}$: Critical path analysis of $\{$Large-Scale$\}$ microservice architectures},
  author={Zhang, Zhizhou and Ramanathan, Murali Krishna and Raj, Prithvi and Parwal, Abhishek and Sherwood, Timothy and Chabbi, Milind},
  booktitle={2022 USENIX Annual Technical Conference (USENIX ATC 22)},
  pages={655--672},
  year={2022}
}

@Article{data-survey-2024-peer-review,
AUTHOR = {Galij, Stanislaw and Pawlak, Grzegorz and Grzyb, Slawomir},
TITLE = {Modeling Data Sovereignty in Public Cloud‚ÄîA Comparison of Existing Solutions},
JOURNAL = {Applied Sciences},
VOLUME = {14},
YEAR = {2024},
NUMBER = {23},
ARTICLE-NUMBER = {10803},
URL = {https://www.mdpi.com/2076-3417/14/23/10803},
ISSN = {2076-3417},
DOI = {10.3390/app142310803}
}

@INPROCEEDINGS{data-sovereignty-2024,
  author={Karagiannis, Vasileios},
  booktitle={2024 11th International Conference on Future Internet of Things and Cloud (FiCloud)}, 
  title={Data Sovereignty and Compliance in the Computing Continuum}, 
  year={2024},
  volume={},
  number={},
  pages={123-130},
  doi={10.1109/FiCloud62933.2024.00027}}

@inproceedings{socc-serverless-carbon,
author = {Sun, Jinghan and Gong, Zibo and Agarwal, Anup and Noghabi, Shadi and Chandra, Ranveer and Snir, Marc and Huang, Jian},
title = {Exploring the Efficiency of Renewable Energy-based Modular Data Centers at Scale},
year = {2024},
isbn = {9798400712869},
publisher = {Association for Computing Machinery},
address = {New York, NY, USA},
url = {https://doi.org/10.1145/3698038.3698544},
doi = {10.1145/3698038.3698544},
booktitle = {Proceedings of the 2024 ACM Symposium on Cloud Computing},
pages = {552–569},
numpages = {18},
location = {Redmond, WA, USA},
series = {SoCC '24}
}

@inproceedings{accountable-carbon-serverless,
author = {Sharma, Prateek and Fuerst, Alexander},
title = {Accountable Carbon Footprints and Energy Profiling For Serverless Functions},
year = {2024},
isbn = {9798400712869},
publisher = {Association for Computing Machinery},
address = {New York, NY, USA},
url = {https://doi.org/10.1145/3698038.3698531},
doi = {10.1145/3698038.3698531},
booktitle = {Proceedings of the 2024 ACM Symposium on Cloud Computing},
pages = {522–541},
numpages = {20},
location = {Redmond, WA, USA},
series = {SoCC '24}
}

@inproceedings{ai-agents,
author = {Shetty, Manish and Chen, Yinfang and Somashekar, Gagan and Ma, Minghua and Simmhan, Yogesh and Zhang, Xuchao and Mace, Jonathan and Vandevoorde, Dax and Las-Casas, Pedro and Gupta, Shachee Mishra and Nath, Suman and Bansal, Chetan and Rajmohan, Saravan},
title = {Building AI Agents for Autonomous Clouds: Challenges and Design Principles},
year = {2024},
isbn = {9798400712869},
publisher = {Association for Computing Machinery},
address = {New York, NY, USA},
url = {https://doi.org/10.1145/3698038.3698525},
doi = {10.1145/3698038.3698525},
booktitle = {Proceedings of the 2024 ACM Symposium on Cloud Computing},
pages = {99–110},
numpages = {12},
location = {Redmond, WA, USA},
series = {SoCC '24}
}

@misc{ms-ai-guide,
  author       = {Zen van Riel},
  title        = {AI System Architecture Essential Guide for Engineers},
  howpublished = {Blog post on zenvanriel.nl},
  year         = {2025},
  month        = {November},
  url          = {https://zenvanriel.nl/ai-engineer-blog/ai-system-architecture-essential-guide-engineers/},
  note         = {Accessed: 2025-11-14}
}

@article{ms-for-large-ai,
  title={Microservices Architecture for Large-Scale AI Applications},
  author={Oye, Emma and Frank, Edwin and Owen, Jane},
  year={2024}
}

@incollection{ms-for-ml,
  title={Facilitating and managing machine learning and data analysis tasks in big data environments using web and microservice technologies},
  author={Shahoud, Shadi and Gunnarsdottir, Sonja and Khalloof, Hatem and Duepmeier, Clemens and Hagenmeyer, Veit},
  booktitle={Transactions on Large-Scale Data-and Knowledge-Centered Systems XLV: Special Issue on Data Management and Knowledge Extraction in Digital Ecosystems},
  pages={132--171},
  year={2020},
  publisher={Springer}
}

@book{ml-for-ms,
  title={Microservices for Machine Learning: Design, implement, and manage high-performance ML systems with microservices (English Edition)},
  author={Ranjan, Rohit},
  year={2024},
  publisher={Bpb Publications}
}

@inproceedings{ml-scale-ms,
author = {Rubak, Adam and Taheri, Javid},
title = {Machine Learning for Predictive Resource Scaling of Microservices on Kubernetes Platforms},
year = {2024},
isbn = {9798400702341},
publisher = {Association for Computing Machinery},
address = {New York, NY, USA},
url = {https://doi.org/10.1145/3603166.3632165},
doi = {10.1145/3603166.3632165},
booktitle = {Proceedings of the IEEE/ACM 16th International Conference on Utility and Cloud Computing},
articleno = {24},
numpages = {8},
location = {Taormina (Messina), Italy},
series = {UCC '23}
}

@ARTICLE{power-modeling-2022,
  author={Radovanovic, Ana and Chen, Bokan and Talukdar, Saurav and Roy, Binz and Duarte, Alexandre and Shahbazi, Mahya},
  journal={IEEE Transactions on Smart Grid}, 
  title={Power Modeling for Effective Datacenter Planning and Compute Management}, 
  year={2022},
  volume={13},
  number={2},
  pages={1611-1621},
  doi={10.1109/TSG.2021.3125275}}

@inproceedings{casper,
author = {Pavlenko, Anna and Cahoon, Joyce and Zhu, Yiwen and Kroth, Brian and Nelson, Michael and Carter, Andrew and Liao, David and Wright, Travis and Camacho-Rodr\'{\i}guez, Jes\'{u}s and Saur, Karla},
title = {Vertically Autoscaling Monolithic Applications with CaaSPER: Scalable Container-as-a-Service Performance Enhanced Resizing Algorithm for the Cloud},
year = {2024},
isbn = {9798400704222},
publisher = {Association for Computing Machinery},
address = {New York, NY, USA},
url = {https://doi.org/10.1145/3626246.3653378},
doi = {10.1145/3626246.3653378},
booktitle = {Companion of the 2024 International Conference on Management of Data},
pages = {241–254},
numpages = {14},
location = {Santiago AA, Chile},
series = {SIGMOD/PODS '24}
}

@inproceedings{ecovisor,
  title={Ecovisor: A virtual energy system for carbon-efficient applications},
  author={Souza, Abel and Bashir, Noman and Murillo, Jorge and Hanafy, Walid and Liang, Qianlin and Irwin, David and Shenoy, Prashant},
  booktitle={Proceedings of the 28th ACM International Conference on Architectural Support for Programming Languages and Operating Systems, Volume 2},
  pages={252--265},
  year={2023}
}

@article{ms-oscillation,
  title={PBScaler: A bottleneck-aware autoscaling framework for microservice-based applications},
  author={Xie, Shuaiyu and Wang, Jian and Li, Bing and Zhang, Zekun and Li, Duantengchuan and Hung, Patrick CK},
  journal={IEEE Transactions on Services Computing},
  volume={17},
  number={2},
  pages={604--616},
  year={2024},
  publisher={IEEE}
}

@inproceedings{imdea-lstms,
author = {Christofidi, Georgia and Papaioannou, Konstantinos and Doudali, Thaleia Dimitra},
title = {Toward Pattern-based Model Selection for Cloud Resource Forecasting},
year = {2023},
isbn = {9798400700842},
publisher = {Association for Computing Machinery},
address = {New York, NY, USA},
url = {https://doi.org/10.1145/3578356.3592588},
doi = {10.1145/3578356.3592588},
booktitle = {Proceedings of the 3rd Workshop on Machine Learning and Systems},
pages = {115–122},
numpages = {8},
location = {Rome, Italy},
series = {EuroMLSys '23}
}

@inproceedings{m-serve-azure,
  title={Power-aware deep learning model serving with $\{$$\mu$-Serve$\}$},
  author={Qiu, Haoran and Mao, Weichao and Patke, Archit and Cui, Shengkun and Jha, Saurabh and Wang, Chen and Franke, Hubertus and Kalbarczyk, Zbigniew and Ba{\c{s}}ar, Tamer and Iyer, Ravishankar K},
  booktitle={2024 USENIX Annual Technical Conference (USENIX ATC 24)},
  pages={75--93},
  year={2024}
}

@inproceedings{usher-azure,
  title={$\{$USHER$\}$: Holistic interference avoidance for resource optimized $\{$ML$\}$ inference},
  author={Shubha, Sudipta Saha and Shen, Haiying and Iyer, Anand},
  booktitle={18th USENIX Symposium on Operating Systems Design and Implementation (OSDI 24)},
  pages={947--964},
  year={2024}
}

@INPROCEEDINGS{microservices-azure,
  author={Nguyen, Hoa X. and Zhu, Shaoshu and Liu, Mingming},
  booktitle={2022 IEEE 11th International Conference on Cloud Networking (CloudNet)}, 
  title={Graph-PHPA: Graph-based Proactive Horizontal Pod Autoscaling for Microservices using LSTM-GNN}, 
  year={2022},
  volume={},
  number={},
  pages={237-241},
  doi={10.1109/CloudNet55617.2022.9978781}}

@ARTICLE{scaling-batch,
  author={Zhang, Qinzhi and Pan, Li and Liu, Shijun},
  journal={IEEE Transactions on Network Science and Engineering}, 
  title={A Cost-Effective Hybrid Cloud Resource Scaling Framework for Batch Processing Services}, 
  year={2025},
  volume={12},
  number={1},
  pages={476-487},
  doi={10.1109/TNSE.2024.3502503}}

@inproceedings{imdea-is-ml-necessary,
author = {Christofidi, Georgia and Papaioannou, Konstantinos and Doudali, Thaleia Dimitra},
title = {Is Machine Learning Necessary for Cloud Resource Usage Forecasting?},
year = {2023},
isbn = {9798400703874},
publisher = {Association for Computing Machinery},
address = {New York, NY, USA},
url = {https://doi.org/10.1145/3620678.3624790},
doi = {10.1145/3620678.3624790},
booktitle = {Proceedings of the 2023 ACM Symposium on Cloud Computing},
pages = {544–554},
numpages = {11},
location = {Santa Cruz, CA, USA},
series = {SoCC '23}
}

@inproceedings{sinan,
  title={Sinan: ML-based and QoS-aware resource management for cloud microservices},
  author={Zhang, Yanqi and Hua, Weizhe and Zhou, Zhuangzhuang and Suh, G Edward and Delimitrou, Christina},
  booktitle={Proceedings of the 26th ACM international conference on architectural support for programming languages and operating systems},
  pages={167--181},
  year={2021}
}

@inproceedings{ms-placement,
author = {Gu, Lin and Zeng, Deze and Hu, Jie and Li, Bo and Jin, Hai},
year = {2021},
month = {05},
pages = {1-9},
title = {Layer Aware Microservice Placement and Request Scheduling at the Edge},
doi = {10.1109/INFOCOM42981.2021.9488779}
}

@misc{electricity-map,
  title = {Electricity Maps },
  howpublished = {\url{https://app.electricitymaps.com/}},
  year = {Accessed: 2025},
}

@misc{api-electricity-map,
  title = {Electricity Maps API},
  howpublished = {\url{https://portal.electricitymaps.com/developer-hub/api/getting-started\#introduction}},
year = {Accessed: 2025}
}

@misc{aws-eks,
  title = {Amazon Elastic Kubernetes Service},
  howpublished = {\url{https://aws.amazon.com/eks/}},
year = {Accessed: 2025}
}

@misc{vpa,
  title = {Kubernetes Vertical Pod Autoscaler},
  howpublished = {\url{https://github.com/kubernetes/autoscaler/tree/master/vertical-pod-autoscaler}},
year = {Accessed: 2025}
}

@misc{liqo,
  title = {Liqo - Enable Dynamic and Seamless Kubernetes Multi-Cluster Topologies},
  howpublished = {\url{https://liqo.io}},
year = {Accessed: 2025}
}

@inproceedings{overcommitment-imdea,
author = {Christofidi, Georgia and Doudali, Thaleia Dimitra},
title = {Do Predictors for Resource Overcommitment Even Predict?},
year = {2024},
isbn = {9798400705410},
publisher = {Association for Computing Machinery},
address = {New York, NY, USA},
url = {https://doi.org/10.1145/3642970.3655838},
doi = {10.1145/3642970.3655838},
booktitle = {Proceedings of the 4th Workshop on Machine Learning and Systems},
pages = {153–160},
numpages = {8},
location = {Athens, Greece},
series = {EuroMLSys '24}
}

@InProceedings{europar-christofidi,
author="Christofidi, Georgia
and Doudali, Thaleia-Dimitra",
editor="Caino-Lores, Silvina
and Zeinalipour, Demetris
and Doudali, Thaleia Dimitra
and Singh, David E.
and Garz{\'o}n, Gracia Ester Mart{\'i}n
and Sousa, Leonel
and Andrade, Diego
and Cucinotta, Tommaso
and D'Ambrosio, Donato
and Diehl, Patrick
and Dolz, Manuel F.
and Jukan, Admela
and Montella, Raffaele
and Nardelli, Matteo
and Garcia-Gasulla, Marta
and Neuwirth, Sarah",
title="Augmenting Cloud Resource Management with¬†the¬†Necessary Amount of¬†Machine Intelligence",
booktitle="Euro-Par 2024: Parallel Processing Workshops",
year="2025",
publisher="Springer Nature Switzerland",
address="Cham",
pages="335--341",
isbn="978-3-031-90203-1"
}

@inproceedings{imdea-midas,
author = {Bang, Tiemo and Roumpos, Ioannis and Matusevych, Sergiy and Christofidi, Georgia and Tian, Yuanyuan and Doudali, Thaleia Dimitra},
title = {Advancing Workload Management with Foundational Models: Challenges in Time Series Similarity and Interpretability},
year = {2025},
isbn = {9798400719608},
publisher = {Association for Computing Machinery},
address = {New York, NY, USA},
url = {https://doi.org/10.1145/3737412.3743491},
doi = {10.1145/3737412.3743491},
booktitle = {Proceedings of the 1st Workshop Connecting Academia and Industry on Modern Integrated Database and AI Systems},
pages = {16–25},
numpages = {10},
location = {
},
series = {MIDAS '25}
}

@misc{lexoffice,
  title = {Case Study: Haufe-Lexware},
  howpublished = {\url{https://aws.amazon.com/solutions/case-studies/haufe-lexware/}},
year = {Accessed: 2025}
}

@misc{aws-latency,
  title = {AWS Latency Monitoring},
  howpublished = {\url{https://www.cloudping.co}},
year = {Accessed: 2025}
}

@misc{aws-pricing,
  title = {AWS Pricing Calculator},
  howpublished = {\url{https://calculator.aws/\#/createCalculator/ec2-enhancement}},
year = {Accessed: 2025}
}

@misc{aws-data-transfer,
  title = {AWS Data Transfer Pricing Calculator},
  howpublished = {\url{https://calculator.aws/\#/createCalculator/DataTransfer}},
year = {Accessed: 2025}
}

@misc{google-regions,
  title = {Global Locations - Regions and Zones, Google Cloud},
  howpublished = {\url{https://cloud.google.com/about/locations}},
year = {Accessed: 2026}
}

@misc{prometheus,
title = {Prometheus Monitoring System},
  howpublished = {\url{https://prometheus.io}},
year = {Accessed: 2025}
}

@misc{n26-1,
title = {Engineering at N26: a Tour of our Tech Stack and Architecture},
  howpublished = {\url{https://medium.com/insiden26/engineering-at-n26-a-tour-of-our-tech-stack-and-architecture-9e58ce96f889}},
year = {2022},
author = {Rafael Roman}
}

@misc{n26-2,
title = {Scaling N26 Technology through Hypergrowth},
  howpublished = {\url{https://www.infoq.com/presentations/scaling-n26/}},
year = {2020},
author = {Folger Fonseca}
}

@misc{ocado,
title = {Ocado Case Study},
  howpublished = {\url{https://kubernetes.io/case-studies/ocado/}},
  year = {Accessed: 2026},
  
}

@misc{docker,
title = {Docker Hardened Images},
  howpublished = {\url{https://hub.docker.com}},
  year = {Accessed: 2026},
}

@misc{cabify,
title = {Cabify},
  howpublished = {\url{https://cabify.com/en}},
  year = {Accessed: 2026}
}

@misc{ittaxi,
title = {itTaxi},
  howpublished = {\url{https://ittaxi.it/en/home-en/}},
  year = {Accessed: 2026}
}

@misc{freenow,
title = {Freenow},
  howpublished = {\url{https://www.free-now.com/es-en/}},
  year = {Accessed: 2026}
}

@article{green,
  title={Green algorithms: quantifying the carbon footprint of computation},
  author={Lannelongue, Lo{\"\i}c and Grealey, Jason and Inouye, Michael},
  journal={Advanced science},
  volume={8},
  number={12},
  pages={2100707},
  year={2021},
  publisher={Wiley Online Library}
}

@misc{aws-ecr,
  title = {Amazon Elastic Container Registry},
  howpublished = {\url{https://aws.amazon.com/ecr/}},
year = {Accessed: 2025},
}

@inproceedings{wang2024autothrottle,
  title={Autothrottle: A Practical $\{$Bi-Level$\}$ Approach to Resource Management for $\{$SLO-Targeted$\}$ Microservices},
  author={Wang, Zibo and Li, Pinghe and Liang, Chieh-Jan Mike and Wu, Feng and Yan, Francis Y},
  booktitle={21st USENIX Symposium on Networked Systems Design and Implementation (NSDI 24)},
  pages={149--165},
  year={2024}
}

@inbook{imbres,
author = {Luo, Shutian and Liao, Jianxiong and Lin, Chenyu and Xu, Huanle and Zhou, Zhi and Xu, Chengzhong},
title = {Embracing Imbalance: Dynamic Load Shifting among Microservice Containers in Shared Clusters},
year = {2025},
isbn = {9798400710797},
publisher = {Association for Computing Machinery},
address = {New York, NY, USA},
url = {https://doi.org/10.1145/3676641.3716255},
booktitle = {Proceedings of the 30th ACM International Conference on Architectural Support for Programming Languages and Operating Systems, Volume 2},
pages = {309–324},
numpages = {16}
}

@INPROCEEDINGS{ursa,
  author={Zhang, Yanqi and Zhou, Zhuangzhuang and Elnikety, Sameh and Delimitrou, Christina},
  booktitle={2024 IEEE International Symposium on High-Performance Computer Architecture (HPCA)}, 
  title={Ursa: Lightweight Resource Management for Cloud-Native Microservices}, 
  year={2024},
  volume={},
  number={},
  pages={954-969},
  doi={10.1109/HPCA57654.2024.00077}}

@article{pause-resume,
author = {Lechowicz, Adam and Christianson, Nicolas and Zuo, Jinhang and Bashir, Noman and Hajiesmaili, Mohammad and Wierman, Adam and Shenoy, Prashant},
title = {The Online Pause and Resume Problem: Optimal Algorithms and An Application to Carbon-Aware Load Shifting},
year = {2023},
issue_date = {December 2023},
publisher = {Association for Computing Machinery},
address = {New York, NY, USA},
volume = {7},
number = {3},
url = {https://doi.org/10.1145/3626776},
doi = {10.1145/3626776},
journal = {Proc. ACM Meas. Anal. Comput. Syst.},
month = dec,
articleno = {45},
numpages = {32}
}

@inproceedings{green-for-less-green,
author = {Hanafy, Walid A and Liang, Qianlin and Bashir, Noman and Souza, Abel and Irwin, David and Shenoy, Prashant},
title = {Going Green for Less Green: Optimizing the Cost of Reducing Cloud Carbon Emissions},
year = {2024},
isbn = {9798400703867},
publisher = {Association for Computing Machinery},
address = {New York, NY, USA},
url = {https://doi.org/10.1145/3620666.3651374},
doi = {10.1145/3620666.3651374},
booktitle = {Proceedings of the 29th ACM International Conference on Architectural Support for Programming Languages and Operating Systems, Volume 3},
pages = {479–496},
numpages = {18},
location = {La Jolla, CA, USA},
series = {ASPLOS '24}
}

@inproceedings{caribou,
  title={Caribou: Fine-Grained Geospatial Shifting of Serverless Applications for Sustainability},
  author={Gsteiger, Viktor Urban and Long, Pin Hong and Sun, Yiran and Javanrood, Parshan and Shahrad, Mohammad},
  booktitle={The 30th ACM Symposium on Operating Systems Principles (SOSP’24). ACM},
  year={2024}
}

@article{carbon-scaler,
author = {Hanafy, Walid A and Liang, Qianlin and Bashir, Noman and Irwin, David and Shenoy, Prashant},
title = {CarbonScaler: Leveraging Cloud Workload Elasticity for Optimizing Carbon-Efficiency},
year = {2023},
issue_date = {December 2023},
publisher = {Association for Computing Machinery},
address = {New York, NY, USA},
volume = {7},
number = {3},
url = {https://doi.org/10.1145/3626788},
doi = {10.1145/3626788},
journal = {Proc. ACM Meas. Anal. Comput. Syst.},
month = dec,
articleno = {57},
numpages = {28}
}

@inproceedings {oakestra,
author = {Giovanni Bartolomeo and Mehdi Yosofie and Simon B{\"a}urle and Oliver Haluszczynski and Nitinder Mohan and J{\"o}rg Ott},
title = {Oakestra: A Lightweight Hierarchical Orchestration Framework for Edge Computing},
booktitle = {2023 USENIX Annual Technical Conference (USENIX ATC 23)},
year = {2023},
isbn = {978-1-939133-35-9},
address = {Boston, MA},
pages = {215--231},
url = {https://www.usenix.org/conference/atc23/presentation/bartolomeo},
publisher = {USENIX Association},
month = jul
}

@inproceedings{dagger,
author = {Lazarev, Nikita and Xiang, Shaojie and Adit, Neil and Zhang, Zhiru and Delimitrou, Christina},
title = {Dagger: efficient and fast RPCs in cloud microservices with near-memory reconfigurable NICs},
year = {2021},
isbn = {9781450383172},
publisher = {Association for Computing Machinery},
address = {New York, NY, USA},
url = {https://doi.org/10.1145/3445814.3446696},
doi = {10.1145/3445814.3446696},
booktitle = {Proceedings of the 26th ACM International Conference on Architectural Support for Programming Languages and Operating Systems},
pages = {36–51},
numpages = {16},
location = {Virtual, USA},
series = {ASPLOS '21}
}

@misc{energy-and-ai-report-2025,
  title = {IEA (2025), Energy and AI, IEA, Paris},
  howpublished = {\url{https://www.iea.org/reports/energy-and-ai}},
license = {CC BY 4.0},
year = {2025}
}

@misc{deathstar-link,
  title = {DeathStar Benchmark},
  howpublished = {\url{https://github.com/delimitrou/DeathStarBench}},
  year = {Accessed: 2025},
}

@inproceedings{DeathStar,
author = {Gan, Yu and Zhang, Yanqi and Cheng, Dailun and Shetty, Ankitha and Rathi, Priyal and Katarki, Nayan and Bruno, Ariana and Hu, Justin and Ritchken, Brian and Jackson, Brendon and others},
title = {An Open-Source Benchmark Suite for Microservices and Their Hardware-Software Implications for Cloud \& Edge Systems},
year = {2019},
isbn = {9781450362405},
publisher = {Association for Computing Machinery},
address = {New York, NY, USA},
url = {https://doi.org/10.1145/3297858.3304013},
doi = {10.1145/3297858.3304013},
booktitle = {Proceedings of the Twenty-Fourth International Conference on Architectural Support for Programming Languages and Operating Systems},
pages = {3–18},
numpages = {16},
location = {Providence, RI, USA},
series = {ASPLOS '19}
}

@article{huo2023hierarchical,
  title={Hierarchical spatio--temporal graph convolutional networks and transformer network for traffic flow forecasting},
  author={Huo, Guangyu and Zhang, Yong and Wang, Boyue and Gao, Junbin and Hu, Yongli and Yin, Baocai},
  journal={IEEE Transactions on Intelligent Transportation Systems},
  volume={24},
  number={4},
  pages={3855--3867},
  year={2023},
  publisher={IEEE}
}

@inproceedings{roy2021sst,
  title={SST-GNN: simplified spatio-temporal traffic forecasting model using graph neural network},
  author={Roy, Amit and Roy, Kashob Kumar and Ahsan Ali, Amin and Amin, M Ashraful and Rahman, AKM Mahbubur},
  booktitle={Pacific-asia conference on knowledge discovery and data mining},
  pages={90--102},
  year={2021},
  organization={Springer}
}

@article{fan2020novel,
  title={A novel model to predict significant wave height based on long short-term memory network},
  author={Fan, Shuntao and Xiao, Nianhao and Dong, Sheng},
  journal={Ocean Engineering},
  volume={205},
  pages={107298},
  year={2020},
  publisher={Elsevier}
}

@inproceedings{shao2022pre,
  title={Pre-training enhanced spatial-temporal graph neural network for multivariate time series forecasting},
  author={Shao, Zezhi and Zhang, Zhao and Wang, Fei and Xu, Yongjun},
  booktitle={Proceedings of the 28th ACM SIGKDD conference on knowledge discovery and data mining},
  pages={1567--1577},
  year={2022}
}

@inproceedings{dote-nsdi,
  title={$\{$DOTE$\}$: Rethinking (predictive)$\{$WAN$\}$ traffic engineering},
  author={Perry, Yarin and Frujeri, Felipe Vieira and Hoch, Chaim and Kandula, Srikanth and Menache, Ishai and Schapira, Michael and Tamar, Aviv},
  booktitle={20th USENIX Symposium on Networked Systems Design and Implementation (NSDI 23)},
  pages={1557--1581},
  year={2023}
}

@article{sibyl,
  title={Sibyl: forecasting time-evolving query workloads},
  author={Huang, Hanxian and Siddiqui, Tarique and Alotaibi, Rana and Curino, Carlo and Leeka, Jyoti and Jindal, Alekh and Zhao, Jishen and Camacho-Rodr{\'\i}guez, Jes{\'u}s and Tian, Yuanyuan},
  journal={Proceedings of the ACM on Management of Data},
  volume={2},
  number={1},
  pages={1--27},
  year={2024},
  publisher={ACM New York, NY, USA}
}

@inproceedings {cilantro,
author = {Romil Bhardwaj and Kirthevasan Kandasamy and Asim Biswal and Wenshuo Guo and Benjamin Hindman and Joseph Gonzalez and Michael Jordan and Ion Stoica},
title = {Cilantro: {Performance-Aware} Resource Allocation for General Objectives via Online Feedback},
booktitle = {17th USENIX Symposium on Operating Systems Design and Implementation (OSDI 23)},
year = {2023},
isbn = {978-1-939133-34-2},
address = {Boston, MA},
pages = {623--643},
url = {https://www.usenix.org/conference/osdi23/presentation/bhardwaj},
publisher = {USENIX Association},
month = jul
}

@inproceedings{take-it-to-the-limit,
author = {Bashir, Noman and Deng, Nan and Rzadca, Krzysztof and Irwin, David and Kodak, Sree and Jnagal, Rohit},
title = {Take It to the Limit: Peak Prediction-Driven Resource Overcommitment in Datacenters},
year = {2021},
isbn = {9781450383349},
publisher = {Association for Computing Machinery},
address = {New York, NY, USA},
url = {https://doi.org/10.1145/3447786.3456259},
doi = {10.1145/3447786.3456259},
booktitle = {Proceedings of the Sixteenth European Conference on Computer Systems},
pages = {556–573},
numpages = {18},
location = {Online Event, United Kingdom},
series = {EuroSys '21}
}

@inproceedings{aquatope,
author = {Zhou, Zhuangzhuang and Zhang, Yanqi and Delimitrou, Christina},
title = {AQUATOPE: QoS-and-Uncertainty-Aware Resource Management for Multi-stage Serverless Workflows},
year = {2022},
isbn = {9781450399159},
publisher = {Association for Computing Machinery},
address = {New York, NY, USA},
url = {https://doi.org/10.1145/3567955.3567960},
doi = {10.1145/3567955.3567960},
booktitle = {Proceedings of the 28th ACM International Conference on Architectural Support for Programming Languages and Operating Systems, Volume 1},
pages = {1–14},
numpages = {14},
location = {Vancouver, BC, Canada},
series = {ASPLOS 2023}
}

@inproceedings{sage,
author = {Gan, Yu and Liang, Mingyu and Dev, Sundar and Lo, David and Delimitrou, Christina},
title = {Sage: practical and scalable ML-driven performance debugging in microservices},
year = {2021},
isbn = {9781450383172},
publisher = {Association for Computing Machinery},
address = {New York, NY, USA},
url = {https://doi.org/10.1145/3445814.3446700},
doi = {10.1145/3445814.3446700},
booktitle = {Proceedings of the 26th ACM International Conference on Architectural Support for Programming Languages and Operating Systems},
pages = {135–151},
numpages = {17},
location = {Virtual, USA},
series = {ASPLOS '21}
}

@article{LiteCast,
  title={LiteCast: A Lightweight Forecaster for Carbon Optimizations},
  author={Joseph, Mathew and Savadi, Tanush and Souza, Abel},
  journal={arXiv preprint arXiv:2511.06187},
  year={2025}
}

@inproceedings{erms,
author = {Luo, Shutian and Xu, Huanle and Ye, Kejiang and Xu, Guoyao and Zhang, Liping and He, Jian and Yang, Guodong and Xu, Chengzhong},
title = {Erms: Efficient Resource Management for Shared Microservices with SLA Guarantees},
year = {2022},
isbn = {9781450399159},
publisher = {Association for Computing Machinery},
address = {New York, NY, USA},
url = {https://doi.org/10.1145/3567955.3567964},
doi = {10.1145/3567955.3567964},
booktitle = {Proceedings of the 28th ACM International Conference on Architectural Support for Programming Languages and Operating Systems, Volume 1},
pages = {62–77},
numpages = {16},
location = {Vancouver, BC, Canada},
series = {ASPLOS 2023}
}

@INPROCEEDINGS{prophet,
  author={Daraghmeh, Mustafa and Agarwal, Anjali and Manzano, Ricardo and Zaman, Marzia},
  booktitle={2021 IEEE International Conference on Communications Workshops (ICC Workshops)}, 
  title={Time Series Forecasting using Facebook Prophet for Cloud Resource Management}, 
  year={2021},
  volume={},
  number={},
  pages={1-6},
  doi={10.1109/ICCWorkshops50388.2021.9473607}}

@inproceedings{how-does-it-function,
author = {Joosen, Artjom and Hassan, Ahmed and Asenov, Martin and Singh, Rajkarn and Darlow, Luke and Wang, Jianfeng and Barker, Adam},
title = {How Does It Function? Characterizing Long-term Trends in Production Serverless Workloads},
year = {2023},
isbn = {9798400703874},
publisher = {Association for Computing Machinery},
address = {New York, NY, USA},
url = {https://doi.org/10.1145/3620678.3624783},
doi = {10.1145/3620678.3624783},
booktitle = {Proceedings of the 2023 ACM Symposium on Cloud Computing},
pages = {443–458},
numpages = {16}, 
location = {, Santa Cruz, CA, USA, }, 
series = {SoCC '23} 
}

@inproceedings{dcperf,
author = {Su, Wei and Dhanotia, Abhishek and Torres, Carlos and Gandhi, Jayneel and Gholkar, Neha and Kanaujia, Shobhit and Naumov, Maxim and Subramanian, Kalyan and Andrei, Valentin and Yuan, Yifan and Tang, Chunqiang},
title = {DCPerf: An Open-Source, Battle-Tested Performance Benchmark Suite for Datacenter Workloads},
year = {2025},
isbn = {9798400712616},
publisher = {Association for Computing Machinery},
address = {New York, NY, USA},
url = {https://doi.org/10.1145/3695053.3731411},
doi = {10.1145/3695053.3731411},
booktitle = {Proceedings of the 52nd Annual International Symposium on Computer Architecture},
pages = {1717–1730},
numpages = {14},
location = {
},
series = {ISCA '25}
}

@INPROCEEDINGS{ecofreq,
  author={Kozlov, Oleksiy M. and Stamatakis, Alexandros},
  booktitle={ISC High Performance 2024 Research Paper Proceedings (39th International Conference)}, 
  title={EcoFreq: Compute with Cheaper, Cleaner Energy via Carbon-Aware Power Scaling}, 
  year={2024},
  volume={},
  number={},
  pages={1-12},
  doi={10.23919/ISC.2024.10528928}}

@inproceedings{seer,
author = {Gan, Yu and Zhang, Yanqi and Hu, Kelvin and Cheng, Dailun and He, Yuan and Pancholi, Meghna and Delimitrou, Christina},
title = {Seer: Leveraging Big Data to Navigate the Complexity of Performance Debugging in Cloud Microservices},
year = {2019},
isbn = {9781450362405},
publisher = {Association for Computing Machinery},
address = {New York, NY, USA},
url = {https://doi.org/10.1145/3297858.3304004},
doi = {10.1145/3297858.3304004},
booktitle = {Proceedings of the Twenty-Fourth International Conference on Architectural Support for Programming Languages and Operating Systems},
pages = {19–33},
numpages = {15},
location = {Providence, RI, USA},
series = {ASPLOS '19}
}

@inproceedings{barista,
  title={Barista: Efficient and scalable serverless serving system for deep learning prediction services},
  author={Bhattacharjee, Anirban and Chhokra, Ajay Dev and Kang, Zhuangwei and Sun, Hongyang and Gokhale, Aniruddha and Karsai, Gabor},
  booktitle={2019 IEEE International Conference on Cloud Engineering (IC2E)},
  pages={23--33},
  year={2019},
  organization={IEEE}
}

@inproceedings {kerveros,
author = {Sultan Mahmud Sajal and Luke Marshall and Beibin Li and Shandan Zhou and Abhisek Pan and Konstantina Mellou and Deepak Narayanan and Timothy Zhu and David Dion and Thomas Moscibroda and Ishai Menache},
title = {Kerveros: Efficient and Scalable Cloud Admission Control},
booktitle = {17th USENIX Symposium on Operating Systems Design and Implementation (OSDI 23)},
year = {2023},
isbn = {978-1-939133-34-2},
address = {Boston, MA},
pages = {227--245},
url = {https://www.usenix.org/conference/osdi23/presentation/sajal},
publisher = {USENIX Association},
month = jul
}

@article{statuscale,
  title={StatuScale: Status-aware and elastic scaling strategy for microservice applications},
  author={Wen, Linfeng and Xu, Minxian and Gill, Sukhpal Singh and Hilman, Muhammad and Srirama, Satish Narayana and Ye, Kejiang and Xu, Chengzhong},
  journal={ACM Transactions on Autonomous and Adaptive Systems},
  volume={20},
  number={1},
  pages={1--25},
  year={2025},
  publisher={ACM New York, NY}
}

@inproceedings{ml-performance-prediction-nsdi,
  title={On the use of $\{$ML$\}$ for blackbox system performance prediction},
  author={Fu, Silvery and Gupta, Saurabh and Mittal, Radhika and Ratnasamy, Sylvia},
  booktitle={18th USENIX Symposium on Networked Systems Design and Implementation (NSDI 21)},
  pages={763--784},
  year={2021}
}

@article{gradient-boosted,
 ISSN = {00905364, 21688966},
 URL = {http://www.jstor.org/stable/2699986},
 author = {Jerome H. Friedman},
 journal = {The Annals of Statistics},
 number = {5},
 pages = {1189--1232},
 publisher = {Institute of Mathematical Statistics},
 title = {Greedy Function Approximation: A Gradient Boosting Machine},
 urldate = {2025-11-14},
 volume = {29},
 year = {2001}
}

@article{random-forests,
  author    = {Breiman, L{eo}},
  title     = {Random Forests},
  journal   = {Machine Learning},
  volume    = {45},
  pages     = {5--32},
  year      = {2001},
  doi       = {10.1023/A:1010933404324}
}

@inproceedings{
patchtst,
title={A Time Series is Worth 64 Words:  Long-term Forecasting with Transformers},
author={Yuqi Nie and Nam H Nguyen and Phanwadee Sinthong and Jayant Kalagnanam},
booktitle={The Eleventh International Conference on Learning Representations },
year={2023},
url={https://openreview.net/forum?id=Jbdc0vTOcol}
}

@book{john2019adaptation,
    author    = {John H. Holland},
    title     = {Adaptation in Natural and Artificial Systems},
    publisher = {MIT Press},
    year      = {1992}
  }

@book{goldberg1989optimization,
  title={Genetic Algorithms in Search, Optimization, and Machine Learning},
  author={Goldberg, David E},
  year={1989},
  publisher={Addison-Wesley}
}

@article{seagull,
author = {Poppe, Olga and Amuneke, Tayo and Banda, Dalitso and De, Aritra and Green, Ari and Knoertzer, Manon and Nosakhare, Ehi and Rajendran, Karthik and Shankargouda, Deepak and Wang, Meina and Au, Alan and Curino, Carlo and Guo, Qun and Jindal, Alekh and Kalhan, Ajay and Oslake, Morgan and Parchani, Sonia and Ramani, Vijay and Sellappan, Raj and Sen, Saikat and Shrotri, Sheetal and Srinivasan, Soundararajan and Xia, Ping and Xu, Shize and Yang, Alicia and Zhu, Yiwen},
title = {Seagull: An Infrastructure for Load Prediction and Optimized Resource Allocation},
year = {2020},
issue_date = {October 2020},
publisher = {VLDB Endowment},
volume = {14},
number = {2},
issn = {2150-8097},
url = {https://doi.org/10.14778/3425879.3425886},
doi = {10.14778/3425879.3425886},
journal = {Proc. VLDB Endow.},
month = {oct},
pages = {154–162},
numpages = {9}
}

@incollection{Thangam2024datacenters,
  author    = {D. Thangam and H. M. and R. Ramesh and G. N. Ramakrishna and N. Muddasir and A. Khan and S. Booshan and B. Booshan and R. Ganesh},
  title     = {Impact of Data Centers on Power Consumption, Climate Change, and Sustainability},
  booktitle = {Computational Intelligence for Green Cloud Computing and Digital Waste Management},
  year      = {2024},
  month     = {mar},
  pages     = {60--83}
}

@inproceedings{alibaba-socc22,
author = {Zhang, Yongkang and Yu, Yinghao and Wang, Wei and Chen, Qiukai and Wu, Jie and Zhang, Zuowei and Zhong, Jiang and Ding, Tianchen and Weng, Qizhen and Yang, Lingyun and Wang, Cheng and He, Jian and Yang, Guodong and Zhang, Liping},
title = {Workload Consolidation in Alibaba Clusters: The Good, the Bad, and the Ugly},
year = {2022},
isbn = {9781450394147},
publisher = {Association for Computing Machinery},
address = {New York, NY, USA},
url = {https://doi.org/10.1145/3542929.3563465},
doi = {10.1145/3542929.3563465},
booktitle = {Proceedings of the 13th Symposium on Cloud Computing},
pages = {210–225},
numpages = {16},
location = {San Francisco, California},
series = {SoCC '22}
}

@inproceedings{cortellessa-ga,
  title={Exploring sustainable alternatives for the deployment of microservices architectures in the cloud},
  author={Cortellessa, Vittorio and Di Pompeo, Daniele and Tucci, Michele},
  booktitle={2024 IEEE 21st International Conference on Software Architecture (ICSA)},
  pages={34--45},
  year={2024},
  organization={IEEE}
}

@inproceedings {cluster-workloads-atc,
author = {George Amvrosiadis and Jun Woo Park and Gregory R. Ganger and Garth A. Gibson and Elisabeth Baseman and Nathan DeBardeleben},
title = {On the diversity of cluster workloads and its impact on research results},
booktitle = {2018 USENIX Annual Technical Conference (USENIX ATC 18)},
year = {2018},
isbn = {978-1-939133-01-4},
address = {Boston, MA},
pages = {533--546},
url = {https://www.usenix.org/conference/atc18/presentation/amvrosiadis},
publisher = {USENIX Association},
month = jul
}

@inproceedings{alibaba21, author = {Luo, Shutian and Xu, Huanle and Lu, Chengzhi and Ye, Kejiang and Xu, Guoyao and Zhang, Liping and Ding, Yu and He, Jian and Xu, Chengzhong}, title = {Characterizing Microservice Dependency and Performance: Alibaba Trace Analysis}, year = {2021}, isbn = {9781450386388}, publisher = {Association for Computing Machinery}, address = {New York, NY, USA}, url = {https://doi.org/10.1145/3472883.3487003}, doi = {10.1145/3472883.3487003}, booktitle = {Proceedings of the ACM Symposium on Cloud Computing}, pages = {412–426}, numpages = {15}, location = {Seattle, WA, USA}, series = {SoCC '21} }

@inproceedings {firm,
author = {Haoran Qiu and Subho S. Banerjee and Saurabh Jha and Zbigniew T. Kalbarczyk and Ravishankar K. Iyer},
title = {{FIRM}: An Intelligent Fine-grained Resource Management Framework for {SLO-Oriented} Microservices},
booktitle = {14th USENIX Symposium on Operating Systems Design and Implementation (OSDI 20)},
year = {2020},
isbn = {978-1-939133-19-9},
pages = {805--825},
url = {https://www.usenix.org/conference/osdi20/presentation/qiu},
publisher = {USENIX Association},
month = nov
}

@inproceedings{amazon-intelligent-scaling,
author = {Diao, Yanlei and Horn, Dominik and Kipf, Andreas and Shchur, Oleksandr and Benito, Ines and Dong, Wenjian and Pagano, Davide and Pfeil, Pascal and Nathan, Vikram and Narayanaswamy, Balakrishnan and Kraska, Tim},
title = {Forecasting Algorithms for Intelligent Resource Scaling: An Experimental Analysis},
year = {2024},
isbn = {9798400712869},
publisher = {Association for Computing Machinery},
address = {New York, NY, USA},
url = {https://doi.org/10.1145/3698038.3698564},
doi = {10.1145/3698038.3698564},
booktitle = {Proceedings of the 2024 ACM Symposium on Cloud Computing},
pages = {126–143},
numpages = {18},
location = {Redmond, WA, USA},
series = {SoCC '24}
}

@incollection{deep-or-stat,
  title={Deep or statistical: an empirical study of traffic predictions on multiple time scales},
  author={Qiao, Yu and Li, Chengxiang and Hao, Shuzheng and Wu, Jun and Zhang, Liang},
  booktitle={Proceedings of the SIGCOMM'22 Poster and Demo Sessions},
  pages={10--12},
  year={2022}
}

@inproceedings{atc-primo-models,
  title={Primo: Practical $\{$Learning-Augmented$\}$ Systems with Interpretable Models},
  author={Hu, Qinghao and Nori, Harsha and Sun, Peng and Wen, Yonggang and Zhang, Tianwei},
  booktitle={2022 USENIX Annual Technical Conference (USENIX ATC 22)},
  pages={519--538},
  year={2022}
}

@inproceedings{random-forest-gpu,
  title={Towards $\{$GPU$\}$ utilization prediction for cloud deep learning},
  author={Yeung, Gingfung and Borowiec, Damian and Friday, Adrian and Harper, Richard and Garraghan, Peter},
  booktitle={12th USENIX Workshop on Hot Topics in Cloud Computing (HotCloud 20)},
  year={2020}
}

@inproceedings{nhits,
  title={Nhits: Neural hierarchical interpolation for time series forecasting},
  author={Challu, Cristian and Olivares, Kin G and Oreshkin, Boris N and Ramirez, Federico Garza and Canseco, Max Mergenthaler and Dubrawski, Artur},
  booktitle={Proceedings of the AAAI conference on artificial intelligence},
  volume={37},
  number={6},
  pages={6989--6997},
  year={2023}
}

\end{document}